\documentclass[aps,prb,superscriptaddress,amsfonts,amsmath,amssymb,twocolumn,showpacs,floatfix]{revtex4}
\usepackage{url}
\usepackage{bm}
\usepackage{graphicx}
\usepackage{amsmath}
\usepackage{amstext}
\usepackage{amssymb}
\usepackage{amsfonts}
\usepackage{amsbsy}
\usepackage{verbatim}
\usepackage{color}
\usepackage[colorlinks=true, urlcolor=blue, linkcolor=blue, citecolor=blue, pdftex]{hyperref}
\usepackage{multirow}

\newcommand{\id}{I}

\include{diagrammatics}

\begin{document}

\title{One-dimensional itinerant interacting non-Abelian anyons}

\author{
\href{http://www.lpt.ups-tlse.fr/spip.php?article32}{Didier Poilblanc}}
\email[]{didier.poilblanc@irsamc.ups-tlse.fr}
\affiliation{Laboratoire de Physique Th\'eorique UMR-5152, CNRS and
Universit\'e de Toulouse, F-31062 France}

\author{Adrian Feiguin}
%\affiliation{Department of Physics and Astronomy,
%University of Wyoming
%Laramie, WY 82071, USA}
\affiliation{Department of Physics, Northeastern
University, Boston, Massachusetts 02115, USA}

\author{Matthias Troyer}
\affiliation{Theoretische Physik, ETH Zurich, 8093 Zurich, Switzerland}

\author{Eddy Ardonne}
\affiliation{%
Nordita, Royal Institute of Technology and Stockholm University,
Roslagstullsbacken 23,
SE-106 91 Stockholm,
Sweden
}
\affiliation{%
Department of Physics, Stockholm University,
AlbaNova University Center, SE-106 91 Stockholm, Sweden
}
\author{Parsa Bonderson}
\affiliation{Station Q, Microsoft Research, Santa Barbara, California 93106-6105, USA}

\date{\today}

\begin{abstract}
We construct models of interacting itinerant non-Abelian anyons moving along one-dimensional chains. We focus on itinerant Ising (Majorana) and Fibonacci anyons, which are, respectively, related to SU$(2)_2$ and SU$(2)_3$ anyons and also, respectively, describe quasiparticles of the Moore-Read and $\mathbb{Z}_3$-Read-Rezayi fractional quantum Hall states.
Following the derivation of the electronic large-U effective Hubbard model,
we derive effective anyonic $t$-$J$ models for the low-energy sectors. Solving these models by exact diagonalization, we find a fractionalization of the anyons into charge and (neutral) anyonic degrees of freedom --  a generalization of spin-charge separation of electrons which occurs in Luttinger liquids. A detailed description of the excitation spectrum can be performed by combining spectra for charge and anyonic sectors.
The anyonic sector is the one of a {\it squeezed} chain of localized interacting anyons, and hence
is described by the same conformal field theory (CFT), with central charge $c=1/2$ for Ising anyons and $c=7/10$ or $c=4/5$ for Fibonacci anyons with antiferromagnetic or ferromagnetic coupling, respectively.
The charge sector is the spectrum of a chain of hardcore bosons subject to
phase shifts which coincide with the momenta of the combined anyonic
eigenstates, revealing a subtle coupling between charge and anyonic excitations at the microscopic level
(which we also find to be present in Luttinger liquids), despite the macroscopic fractionalization.
The combined central charge extracted from the entanglement entropy between segments of the chain is shown to be $1+c$, where $c$ is the central charge of the underlying CFT of the localized anyon (squeezed) chain.
\end{abstract}

\maketitle

\section{Introduction}

One of the most significant pursuits in condensed matter physics is the search for quasiparticles or excitations that obey non-Abelian exchange statistics~\cite{Leinaas77,Goldin85,Fredenhagen89,Froehlich90}. The most prominent candidates (at present) are excitations of non-Abelian quantum Hall states~\cite{Moore91,Read99,Lee07,Levin07,Bonderson07d}. In particular, there is evidence from tunneling~\cite{Radu08} and interferometry~\cite{Willett09a,Willett12} experiments supporting the existence of non-Abelian quasiparticle excitations for the $\nu=5/2$ quantum Hall state~\cite{Willett87,Pan99,Eisenstein02}.

The leading candidate quantum Hall states to describe the electronic ground state
of the quantum Hall plateau at filling fraction $\nu=5/2$ are the Moore-Read (MR) Pfaffian
state~\cite{Moore91} or its particle-hole conjugate, the ``anti-Pfaffian'' (aPf) state~\cite{Lee07,Levin07}.
Despite the differences between the MR and aPf states
(which manifest in the detailed structure of the edge states),
the non-Abelian anyonic structure of their bulk quasiparticles are simply the
complex conjugates of each other. Both can
be described in terms of an Ising-type anyon model.

One of the leading candidates to describe the experimentally observed $\nu = 12/5$ quantum Hall
plateau~\cite{Xia04,Kumar10} is the $k=3$ Read-Rezayi (RR) state~\cite{Read99} (a generalization of the MR
state), or, more precisely, its particle-hole
conjugate ($\overline{\text{RR}}$). The non-Abelian quasiparticles of the RR and $\overline{\text{RR}}$ states are of
Fibonacci type, and the $\nu=12/5$ quantum Hall state is the leading candidate system
hosting such non-Abelian anyons. The other leading candidate for describing the $\nu=12/5$ quantum Hall effect is provided by
Bonderson-Slingerland (BS) states~\cite{Bonderson07d} obtained hierarchically from the MR and aPf $\nu=5/2$ states (by condensing Laughlin-type quasiholes).
The quasiparticles of these BS states have a similar Ising-type non-Abelian structure as their MR and aPf parent states.
Numerical studies of the $\nu=12/5$ quantum Hall state found the $\overline{\text{RR}}$ and BS candidates to be in close competition~\cite{Bonderson09a}.

Interestingly, a different hierarchical construction over the MR state (condensing fundamental non-Abelian quasielectrons) produces a candidate state for filling $\nu = 18/7$ that possesses non-Abelian quasiparticles of the Fibonacci type~\cite{h10}, similar to the ones appearing in the non-Abelian spin-singlet (NASS) state~\cite{Ardonne99}. However, a quantum Hall state at $\nu = 18/7$ has, so far, not been experimentally observed.

Another promising class of candidates for realizing non-Abelian quasiparticles is provided by systems with the so-called emergent Majorana zero modes, which behave like Ising-type anyons under exchange. Majorana zero modes were originally predicted to exist in vortex cores of chiral p-wave superconductors~\cite{Volovik99,Read00} or at the ends of one-dimensional polarized superconductors~\cite{Kitaev01a}.
More recently, it was shown that Majorana fermions can form at the interface of a strong topological insulator and an
s-wave superconductor~\cite{fk08}. This idea for realizing Majorana fermion zero modes was further developed by several groups~\cite{slt10,a10,Lutchyn10,Oreg10},
who proposed similar superconducting heterostructures based on semiconductors exhibiting strong spin-orbit coupling, rather than topological insulators. For a review, see Ref.~\onlinecite{a12}. Efforts to physically implement these latest proposals have been made in recent experiments~\cite{mzf12,Rokhinson12,dyh12up,Das12}
consisting of electrons tunneling into a nanowire (set-up to be in a topological phase supporting edge Majorana modes).
However, neither the exponential localization to the edges of the observed zero-modes nor the non-Abelian exchange statistics
has been probed yet.

Inspired by these recent developments and possible realizations of quasiparticles with non-Abelian statistics, we consider the question of what happens
if one confines mobile non-Abelian quasiparticles to one-dimensional (1D) systems.
The concept of itinerancy of interacting non-Abelian quasiparticles is of direct physical significance,
and the microscopic models we study can be viewed, for example, as (crude) effective
models relevant to edge modes of quantum Hall and Majorana fermion systems.
It was established long ago, starting with the work of Anderson~\cite{anderson}, that electrons confined to one dimension
undergo ``spin-charge separation,'' namely the electrons falls apart into two pieces, one
spinless carrying the charge, the other a spinon without charge, carrying the spin. These
ideas were further developed by several people, Tomonaga~\cite{t50}, Luttinger~\cite{l63},
and Haldane~\cite{h81}, who introduced the concept of the one-dimensional
{\em Luttinger liquid}.

In our recent Letter~\cite{short-2012}, we started to investigate the subject of itinerant non-Abelian anyons in a one-dimensional system. We established
that non-Abelian anyons (of which the quasiparticles of the quantum Hall states and the Majorana zero modes discussed above are prime
examples) also undergo a process which resembles spin-charge separation. Namely,
the non-Abelian anyons fractionalize into charge (or density) and anyonic degrees of
freedom. The model introduced in Ref.~\onlinecite{short-2012} was inspired by the electronic $t$-$J$
model~\cite{ZhangRice}, which can be viewed as a limiting case of the Hubbard model~\cite{Hubbard1,Hubbard2,Hubbard3} -- namely in the limit of large on-site repulsion -- and for which spin-charge separation was established analytically at a supersymmetric point~\cite{bb90,tJ-1D-kawakami,bbo91}
and numerically~\cite{tJ-1D}. In this paper, we continue our study of itinerant non-Abelian anyons and provide greater detail.

The outline of the paper is as follows: in Sec.~II we review briefly general properties of
non-Abelian anyons in SU$(2)_k$ Chern-Simons theories and, more specifically, in non-Abelian
quantum Hall states. In Sec.~III, we show that, in close analogy to the electronic case, it is possible to: (i) truncate the Hilbert space of the quasiparticles of the non-Abelian quantum Hall states confined to a one-dimensional geometry (in the case of strong charging energy), and (ii) derive low-energy effective anyonic $t$-$J$ models.
The charge sectors of the anyonic $t$-$J$ models are derived in Sec.~IV.
In Sec.~V, we present a short review of the properties of dense (localized)
non-Abelian anyon chains. In Sec.~VI, we demonstrate that
the excitation spectrum of the anyonic $t$-$J$ models
can be accurately described by combining spectra for charge and anyonic sectors (in a subtle manner), and provide clear evidence of the fractionalization of anyons into
charge and anyonic degrees of freedom. In Sec.~VII, we use density matrix renormalization group (DMRG) calculations to extract the central charge of the Fibonacci
$t$-$J$ chain from the entanglement entropy between segments of open chains, providing further evidence of fractionalization. In Appendix A, we provide detailed descriptions of the related anyon
models of the MR and RR states. In Appendix~B, we describe the details of the quasiparticle spectrum truncation for MR and RR anyons used in the paper.

\section{Non-Abelian anyons}

\subsection{General considerations and fusion algebra}

The non-Abelian anyons we will be concerned with in this
paper can formally be described by SU$(2)_k$ Chern-Simons theories,
or via a certain quantum deformation of SU$(2)$. In either case, the
non-Abelian degrees of freedoms are captured by
the ``topological charges'' $j$, which can be thought of as ``generalized angular momenta.'' For a given SU(2)$_k$ theory,
these are constrained to take the values $j=0,\frac{1}{2},\ldots, \frac{k}{2}$, loosely
corresponding to the first $k+1$ representations of SU$(2)$.

In the same way that the tensor product of SU$(2)$ spins can be decomposed
into the direct sum of multiplets of definite values of $J^2$, one can decompose the product, or ``fusion'' of anyons.
This fusion algebra or ``fusion rules'' of a general anyon model takes the form
\begin{equation}
a \times b = \sum_{c \in \mathcal{C}} N^{c}_{ab}\, c
\end{equation}
where $a$, $b$, and $c$ are topological charge values in the set of allowed topological charges $\mathcal{C}$, and the fusion coefficients $N^{c}_{ab}$ are non-negative integers indicating the number of ways $a$ and $b$ can fuse to produce $c$. The $N^{c}_{ab}$ must be such that the algebra is commutative and associative. There must also be a unique ``vacuum'' or ``trivial'' charge, which we denote as $I$ or $0$, for which $N^{c}_{a0}=\delta_{ac}$.

The fusion rules of SU(2)$_k$ anyons resemble their SU$(2)$ counterpart (but with a finite set of allowed values for $j$ and a corresponding truncation of the algebra). In particular
\begin{equation}
j_1 \times j_2 = \sum_{j_3 = |j_1-j_2|}^{\min\{ j_1+j_2 , k-j_1-j_2 \}} j_3 \ ,
\end{equation}
where the upper limit is such that the fusion rules are associative and
obey the constraint that $j_i \leq \frac{k}{2}$.

We will be particularly interested in the cases $k=2$ and $3$, because they are
the experimentally most relevant non-Abelian anyon models.
(The case $k=1$ corresponds to Abelian anyon models, in which case the Hilbert spaces we construct below are one-dimensional, so one can not construct non-trivial models.)
These cases are related to the Ising and Fibonacci anyons models, respectively, which we now consider in more detail.

The anyon model with $k=2$ has three anyon types, which we will label using the Ising TQFT topological charges $I$, $\sigma$, and $\psi$
as follows: The vacuum or trivial anyon $I$ corresponds to $j=0$. The anyon of type $\sigma$ corresponds to
$j=\frac{1}{2}$. Finally, the (fermionic) anyon type $\psi$ corresponds to $j=1$.
In particular, the fusion rules read
\footnote{The Ising TQFT has the same fusion algebra as SU$(2)_2$, but the $\sigma$ and $j=\frac{1}{2}$ anyons have different scaling dimensions ($h_{\sigma}=1/16$ and $h_{\frac{1}{2}}=3/16$, respectively. The scaling dimensions of the $\psi$ and $j=1$ anyons are identical, $h_{\psi} = h_{1} = 1/2$).}
\begin{align}
\sigma \times \sigma &= I+\psi & \sigma \times \psi &= \sigma & \psi \times \psi = I \ ,
\end{align}
in addition to the general relations $\alpha \times \beta = \beta \times \alpha$
and $I \times \alpha = \alpha$, which hold in all anyon models, for arbitrary $\alpha$ and $\beta$.

The Fibonacci anyon model correspond to the case $k=3$, where we restrict ourselves
to the integer-valued $j=0$ and $1$, which we will label as the vacuum $I$ and the Fibonacci anyon $\tau$,
respectively.
\footnote{
For $k$ odd, the restriction to the integer-valued topological charge (generalized angular momentum) can be made by using the
map obtained by fusing with the topological charge $j=\frac{k}{2}$, namely $j \times \frac{k}{2} = \frac{k}{2} - j$,
which for $k=3$ maps $\frac{1}{2} \leftrightarrow 1$ and $\frac{3}{2} \leftrightarrow 0$. For more details,
we refer to Refs.~\onlinecite{Bonderson07b,Trebst08a}.
}
The Fibonacci fusion rule reads
\begin{equation}
\tau \times \tau = I + \tau \ .
\end{equation}

\begin{figure}
\includegraphics[width=0.9\columnwidth]{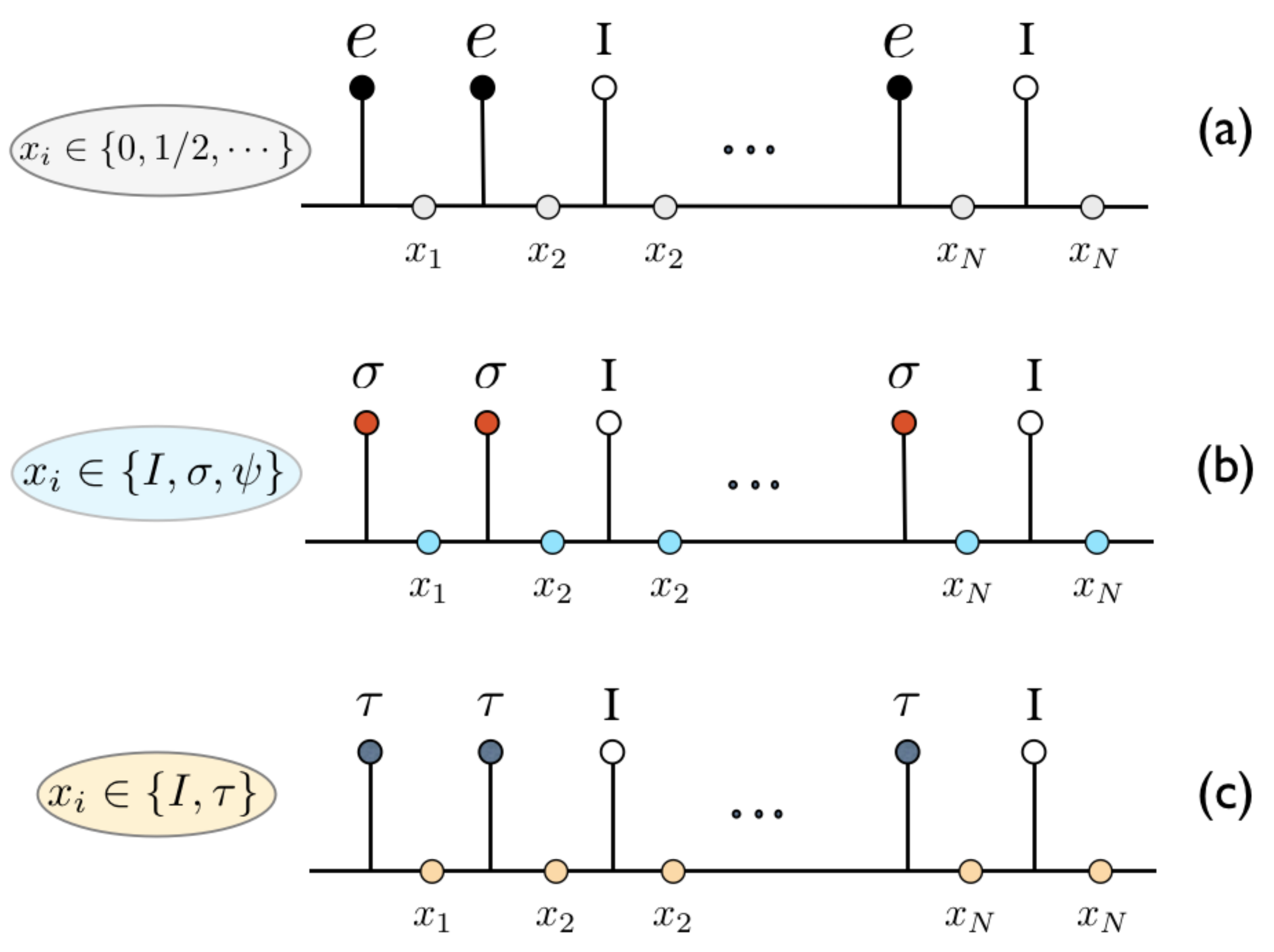}
\caption{Anyonic fusion trees for (a) electrons, (b) Ising $\sigma$ anyons, and (c) Fibonacci
anyons. Sites are denoted by the (filled or empty) circles at top of the diagrams. Empty circles denote
vacant sites, which carry the vacuum or trivial topological charge $0$ or $I$.
The bond labels $\{x_i\}$ encode non-local information about the state and their possible values are specified for each model.
We note that, for Ising anyons, our model excludes $\psi$ anyons on the sites, but not on the links (see the main text).}
\label{fig:sketch}
\end{figure}

In the following we shall consider itinerant anyons moving on one dimensional chains.
Pictorial representations of such anyonic
chains are shown in Fig.~\ref{fig:sketch}(b,c), together with the more familiar
case of strongly correlated electrons shown in Fig.~\ref{fig:sketch}(a).

For the case of strongly correlated electrons, each electron carries a unit charge and spin-$\frac{1}{2}$. The spin-$\frac{1}{2}$ degrees of freedom are taken
into account in Fig.~\ref{fig:sketch}(a) in a slightly unconventional way, utilizing a ``fusion tree'' notation, rather than the usual tensor product of $N$ two-dimensional local Hilbert spaces (where $N$ is the number of electrons). In this notation, the labels $x_i$ for the links of the fusion tree correspond to the total spin obtained by combining the spin $x_{i-1}$ with that of the $i$th electron. For an open chain, this simply means that $x_i$ is the total spin of all the electrons to the left of the label. For a periodic chain (in a system on a torus), the label has a slightly more abstract interpretation, since the notion of all particles to the left or right are not well-defined. We use this formulation because it easily generalizes to the case of non-Ablian anyons, where there are no local degrees of freedom (i.e. they lack local Hilbert spaces and internal quantum numbers, similar to $s_z$ in the case of spins).

The non-Abelian anyons in Fig.~\ref{fig:sketch}(b,c) may also carry electric charge
(albeit this typically is a fraction of the charge of the electron), as well as
anyonic degrees of freedom. The charge degrees of freedom
live on the sites, while the bond variables $x_i$  encode the
anyonic degrees of freedom along the fusion tree, in the same way as
the labels $x_i$ encoded the spin of the electrons in Fig.~\ref{fig:sketch}(a). Abelian anyonic degrees of freedom may be
treated in the same way as electric charge, i.e. locally assigned to the sites, since their resulting fusion tree is uniquely determined by the local degrees of freedom.
The labels $x_i$ are not arbitrary, but satisfy the constraint that each trivalent
vertex in this fusion tree is permitted by the fusion rules. This implies that
the size of the {\em internal} Hilbert space (for a given configuration of particle/anyon positions)
grows as $2^N$ in the case of the electrons,
$(\sqrt{2})^N$ in the case of the Ising anyons,
and $\phi^N$ in the case of the Fibonacci anyons, where $\phi = \frac{1+\sqrt{5}}{2}$
is the golden ratio. Here, $N$ corresponds
to the number of electrons, Ising $\sigma$ anyons, or Fibonacci anyons. The actual dimension for
any finite $N$ is, of course, an integer, so these are only the leading order scaling (as $N \rightarrow \infty$) for the non-Abelian anyons.
The sites labeled by $I$ correspond to vacancies,
and carry no electric charge, spin, or anyonic degrees of freedom.

Before we continue in the next subsection with describing the quantum Hall states
in which these types of anyons are realized, we want to make one remark, which
will be essential in the subsequent description of the behavior of the itinerant anyons.
Despite the fact that we will be describing mobile, but identical, anyons, there
will be a notion of ``distinguishability'' of the anyons. In particular, the various states in the
Hilbert space are not only characterized by the location of the occupied sites, but also
by the labels $x_i$, which distinguish the various states, given the location of all the anyons.
In some sense, specifying the precise internal state, corresponding to all the anyons as
a whole, renders the individual anyons in a particular state distinguishable. We will see
later on that this seemingly simple observation will play an essential role in the effective
description of the collective behavior of the itinerant anyons.

\subsection{Non-Abelian quantum Hall states}
\label{subsec:MRRR}

We now concentrate on describing the anyonic structure of the MR and
$k=3$ RR states in their fermionic incarnations, which are relevant in the
electronic quantum Hall setting. Because of the fermionic nature of the states, the
anyonic structure is slightly more complicated than the SU$(2)_k$ anyons described
above. To describe this structure, it is best to consider the non-Abelian part separately,
which is described in terms of Ising anyons for the MR state, and $\mathbb{Z}_3$
parafermions for the $k=3$ RR state.

\begin{figure}
\includegraphics[width=0.8\columnwidth]{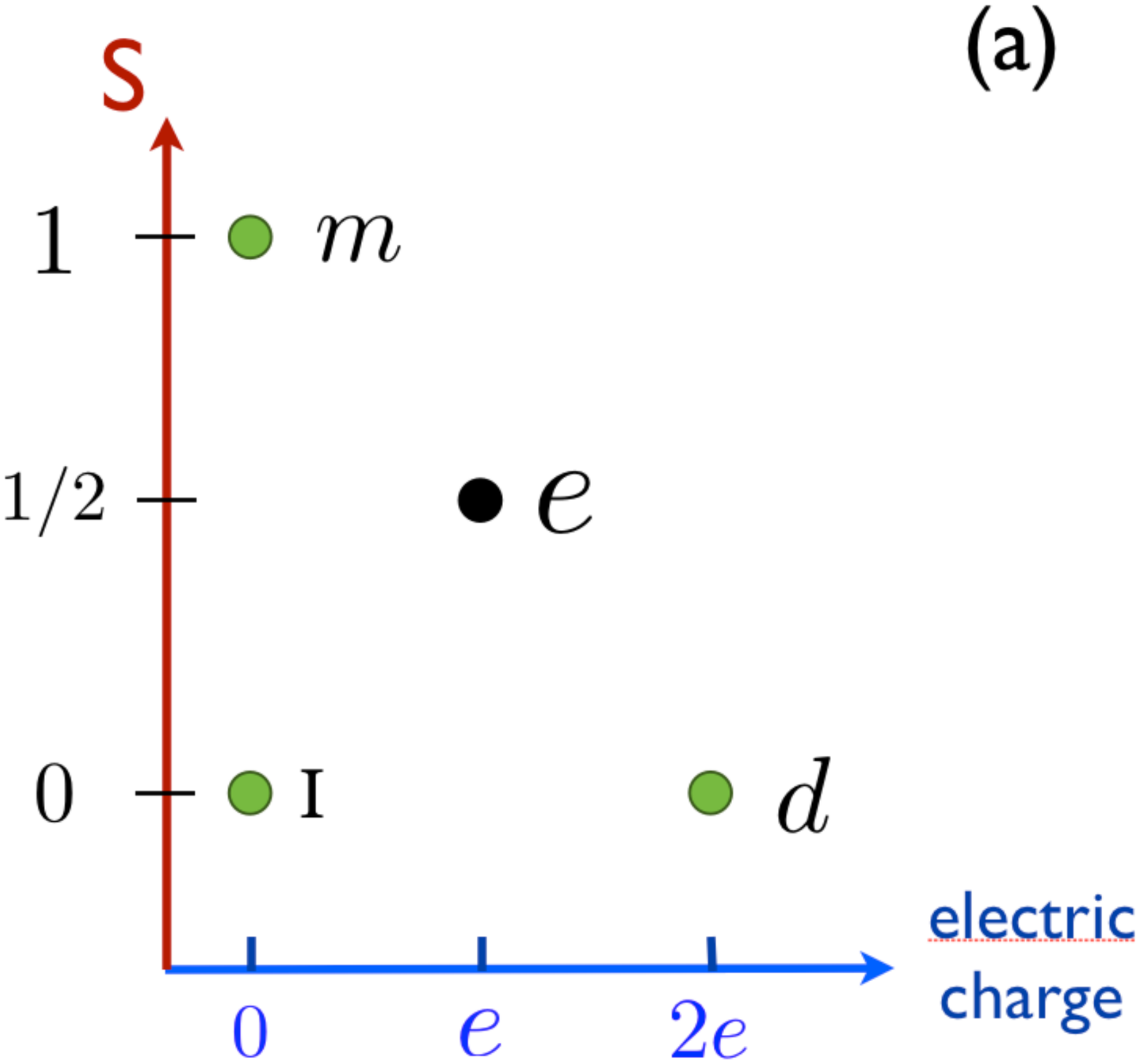}
\vskip 0.25truecm
\includegraphics[width=0.8\columnwidth]{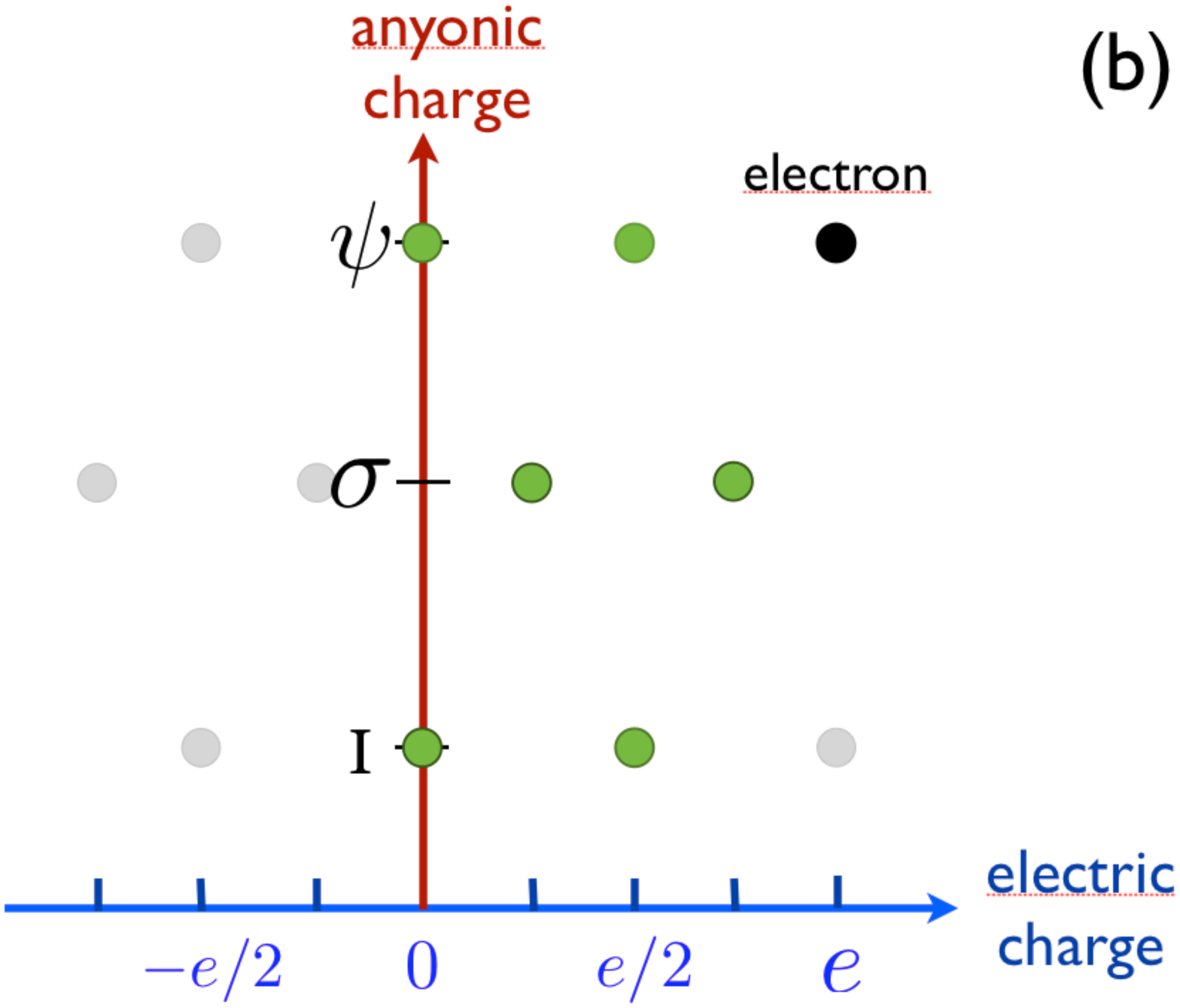}
\vskip 0.25truecm
\includegraphics[width=0.8\columnwidth]{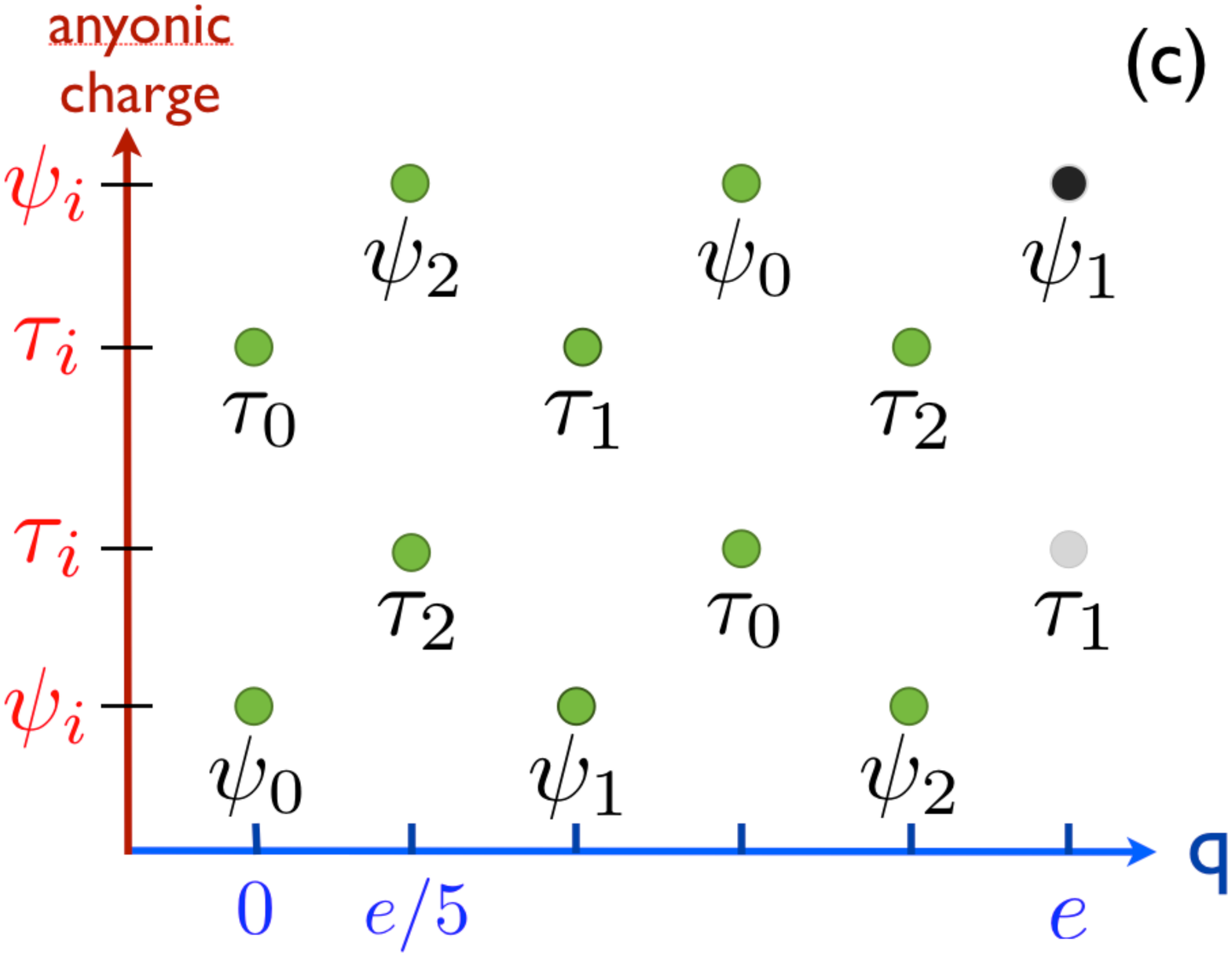}
\caption{Charts of the quasiparticle contents of (a) the electronic Hubbard model, (b) the MR state and (c) the $k=3$ RR state.
The elementary electric charges of the MR quasiparticles are multiples of $e/4$, while that of the RR quasiparticles are multiples of $e/5$.
The dark (green) symbols correspond to the different particle types. The black dots represents electrons/holes, which are identified with the vacuum $(I,0)$ in (b) and the vacuum $(\psi_0,0)$ in (c). The grey dots correspond to particles which are identified with one of
particles corresponding to a green symbol, as explained in the main text.
%In (b) antiparticles are shown as grey dots. The electron of charge $-|e|$ can be identified with
%the ``positron" of charge $+|e|$.
}
\label{fig:charts}
\end{figure}

In the case of the MR state, the non-Abelian sector is the Ising theory, whose
fields are $I, \sigma, \psi$, with the Ising fusion rules given above.
The quasiparticle types can now be specified by the Ising label, together with the
electric charge. The vacuum is $(I,0)$, while $(\sigma,e/4)$ is the ``fundamental quasiparticle,'' which, in some sense, carries the ``smallest'' quantum numbers allowed in the MR state, i.e. it has the smallest (nonzero) electric charge and repeated fusion generates the entire spectrum of topological charges.
All other quasiparticles are thus obtained by repeated fusion of this fundamental quasiparticle,
using the fusion rules above and the additivity of the charge. In addition, one needs the
rule that quasiparticles which differ by fusion of an electron, given by $(\psi,e)$ are to be identified.
The fact that we identify quasiparticles which ``differ by a fermion'' (or identify the electron with
the vacuum) leads to some complications, which are not present for the bosonic versions
of these quantum Hall states~\footnote{%
For the bosonic states, the particle identified with the vacuum is a boson.
In the fermionic case, one cannot simply identify the electron with the vacuum, because
it is a fermion, which obviously has different braiding statistics than the vacuum
(which is a boson). Considering fusion and braiding,
one could instead simply identify pairs of electrons with vacuum.
However, the resulting theory will not be modular, meaning the S-matrix
is degenerate. This poses a problem when one wishes to define the theory
on arbitrary surfaces, including the torus. A solution is to put each charge into
a $Z_2$ doublet, e.g. the vacuum and electron form the vacuum doublet,
and every charge together with the charge obtained by fusion with an electron
form a doublet. Then the S-matrix of doublets is modular. In practice, one can
take the fusion rules assuming identification of electron with vacuum.
},
but these complications will not concern us here.
The resulting quasiparticle spectrum is given in Fig.~\ref{fig:charts}(b), where we have six different
quasiparticle types (shown as green circles), because $(I,e)$ and $(\psi,0)$ are identified, and so on.

In the case of the RR state with $k=3$, the non-Abelian structure corresponds
to the $\mathbb{Z}_3$-parafermions, which conventionally are labeled as
$I,\psi_{1},\psi_{2},\sigma_1,\sigma_2,\varepsilon$. We will, however, use the notation
$\psi_0 = I$, $\tau_0 = \varepsilon$, $\tau_2 = \sigma_1$
and $\tau_1 = \sigma_2$. In this way, the fusion rules take the simple form
\begin{align}
\psi_i \times \psi_j &= \psi_{i+j} \nonumber \\
\psi_i \times \tau_j &= \tau_{i+j} \\
\tau_i \times \tau_j &= \psi_{i+j} + \tau_{i+j} \ ,\nonumber
\end{align}
where the indices are taken modulo 3. This is the direct product of the Fibonacci fusion algebra with a $\mathbb{Z}_3$ fusion algebra.
The quasiparticle types can now be specified by the $\mathbb{Z}_3$-parafermion label, together with the
electric charge. The vacuum is $(\psi_0,0)$ while the fundamental quasiparticle's quantum numbers
are $(\tau_2,e/5)$. As in the MR state, the other quasiparticles are obtained by repeated
fusion of the fundamental quasiparticle, using the fusion rules above. Quasiparticles which differ by
fusion of an electron, given by $(\psi_1,e)$, are to be identified. This gives rise to ten different
quasiparticle types displayed as green circles in Fig.~\ref{fig:charts}(c), with charges $0,\ldots, 4e/5$ [where we note that $(\tau_1,e)$ is identified with $(\tau_0,0)$, and so on].

For comparison, we display the relevant quasiparticle types in the ordinary electron case in Fig.~\ref{fig:charts}(a): the trivial particle (vacant site) $I$, the electron $e$, a double occupied site $d$ and a spin-1 magnon $m$. In this case, there is no condensate, and hence none of the particles are to be identified.

In the next section, we will describe how we can truncate the spectrum of particles, in order
to come up with a tractable model of interacting, itinerant anyons in one dimension.

\section{Anyonic $t$-$J$ models}

\subsection{Objectives and procedure}

\begin{figure}
\includegraphics[width=0.8\columnwidth]{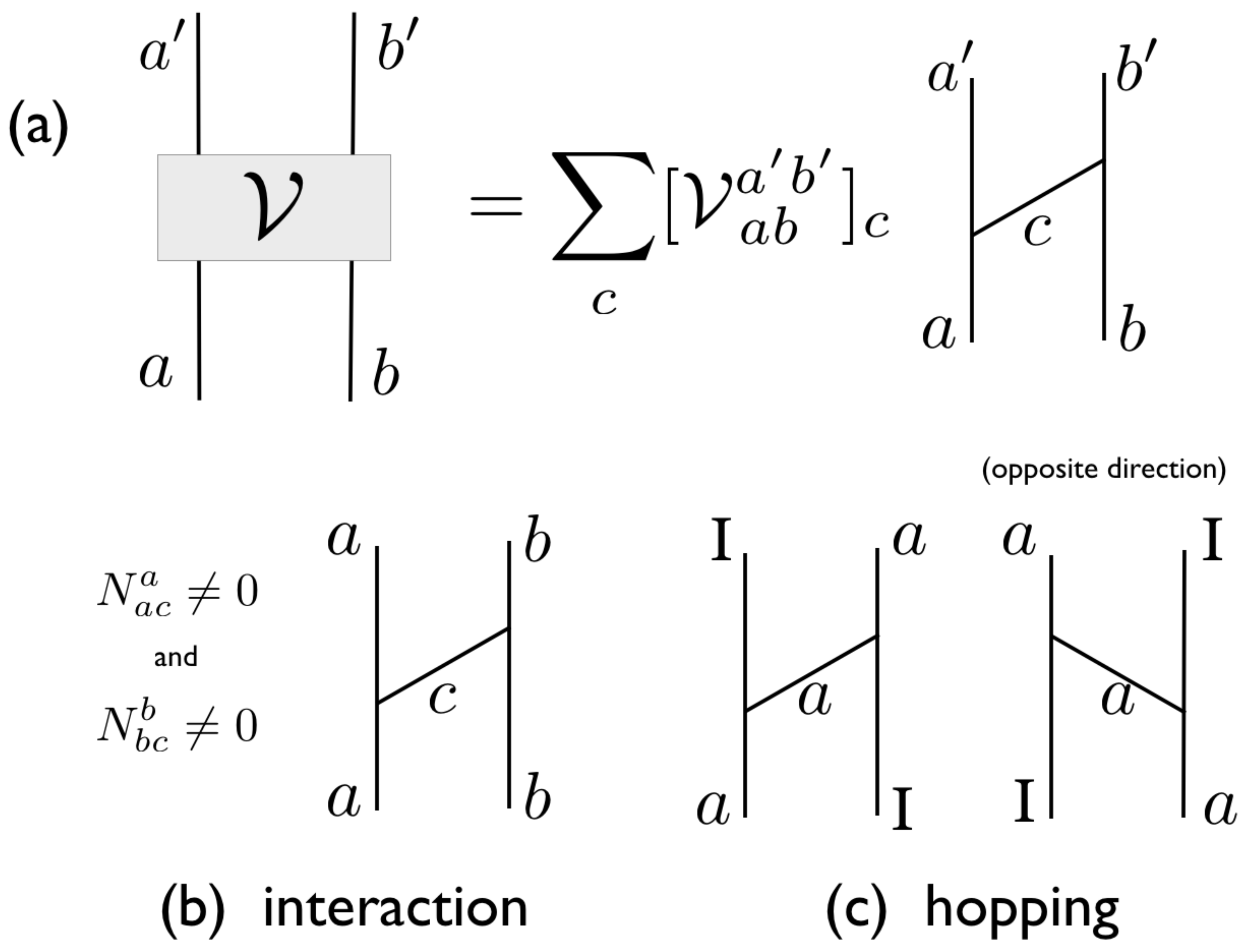}
\caption{(a) All two-anyon processes can be represented as a general ``tunneling'' term, where topological (and/or electrical) charge is transferred from one localized quasiparticle to the other. Special cases of this include (b) an ``interaction'' between the two anyons, for which the localized charges are unchanged, and (c) a ``hopping'' term, where a localized anyon moves to a vacant site.}
\label{fig:interactions}
\end{figure}

\begin{figure}
\includegraphics[width=0.8\columnwidth]{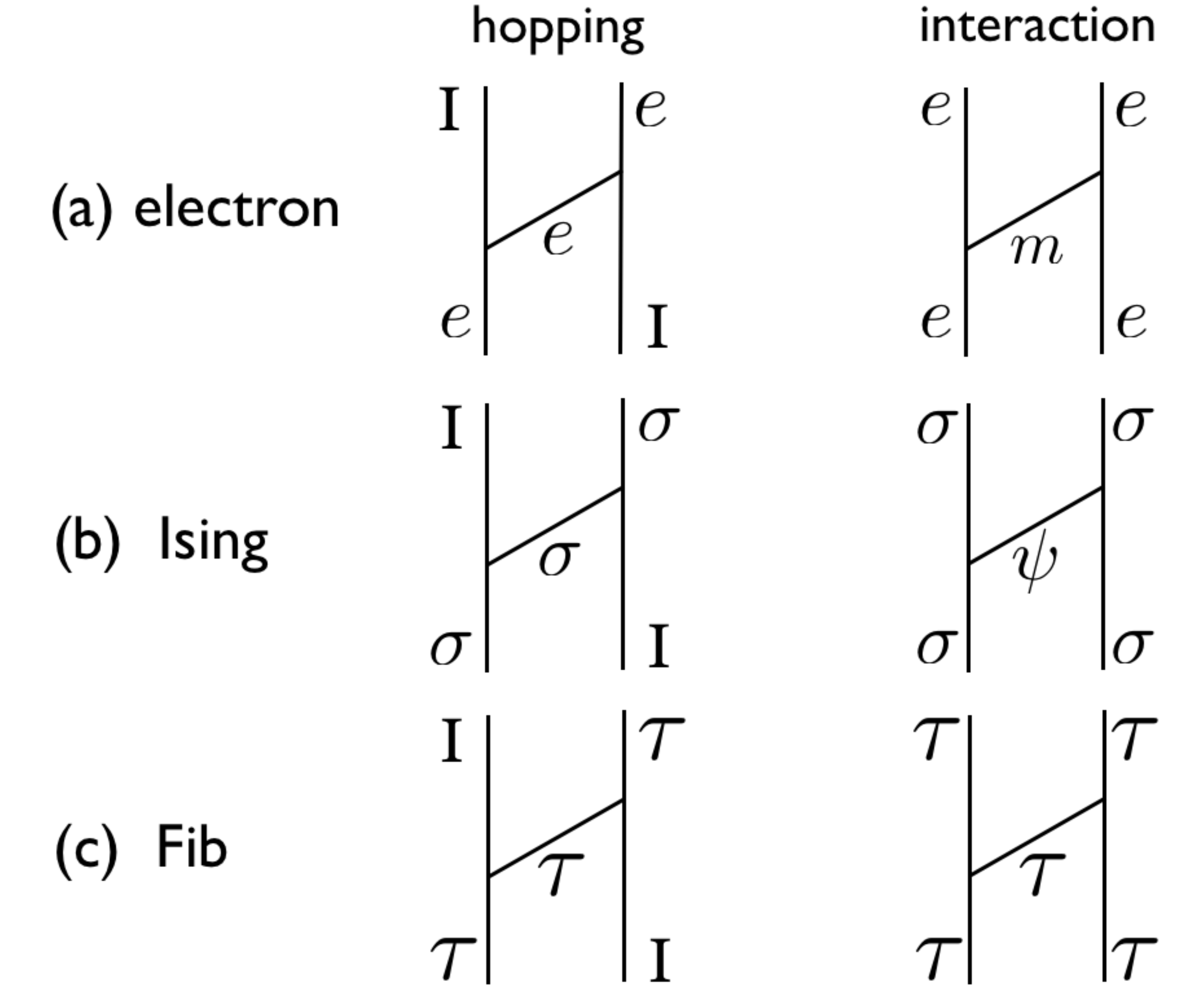}
\caption{Hopping and interaction terms for (a) electrons, (b) Ising anyons, and (c) Fibonacci anyons, expressed in the notations
of Fig.~\ref{fig:interactions}. }
\label{fig:interactions2}
\end{figure}

We move now to the construction of the low energy models for the itinerant anyons in one dimension, modeled by a discrete chain. This chain might be a lattice discretization of a one-dimensional continuum system such as the edge of a quantum Hall liquid or a one-dimensional array of quantum dots. On this chain we restrict ourselves to short-range (nearest-neighbor) interactions and ``hopping'' terms that can both be expressed as ``tunneling'' processes~\cite{Bonderson09b}, as sketched in Figs.~\ref{fig:interactions},\ref{fig:interactions2}.
Because the anyons are electrically charged, confinement on a quantum dot or transverse confinement (e.g. in the case of edge states)
may lead to large Coulombic charging energy that strongly discourages multiple occupancy of sites and, hence, prohibits the formation of quasiparticle excitations of larger charge values.

As seen in the previous section the physical contents of the non-Abelian quantum Hall states are very rich and we thus want to derive an effective low-energy model, similar to the derivation of the $t$-$J$ model for electrons from the Hubbard model.
In order to describe the low-energy spectra, a well-know strategy consists of
building up a simpler model by (i) discarding the high-energy (quasi)particles
and (ii) treating virtual processes involving the fusion of the low-energy (quasi)particles to the discarded high-energy (quasi)particles in second-order perturbation theory.

\subsection{Large-U electronic Hubbard model}

To illustrate this procedure, we first take the example of a generalized Hubbard model of electrons and show
how to derive the corresponding $t$-$J$ model.
We start from electrons, where the most general SU$(2)$-symmetric single band model with nearest neighbor interactions can be  written in second quantized notation as
\begin{eqnarray}
H&=&-t\sum_{i,\sigma}\left(c^\dag_{i,\sigma}c^{\vphantom\dag}_{i+1,\sigma} + H.c.\right) \nonumber \\
&& +J_0\sum_i\left(\vec{S}_i\cdot\vec{S}_{i+1}-\frac{1}{4}n_{i}n_{i+1}\right) \nonumber \\
&& +V\sum_in_in_{i+1}\nonumber \\
&&+U\sum_i n_{i,\uparrow}n_{i,\downarrow}.
\label{eq:H_t_J}
\end{eqnarray}
Here $c^\dag_{i,\sigma}$ and $c_{i,\sigma} $ are the creation an annihilation operators of an electron with $z$-component of spin $\sigma$, $n_{i,\sigma} = c^\dag_{i,\sigma}c^{\vphantom\dag}_{i,\sigma}$ are the local spin densities, $n_i=n_{i,\uparrow}+n_{i,\downarrow}$ is the total local density, and $\vec{S}_i$ is the spin operator on site $i$.

The first term ($t$) in the Hamiltonian of Eq.~(\ref{eq:H_t_J}) is the hopping (tunneling) of an electron.
The second term ($J_0$) is a spin exchange term, which can be interpreted as a two-electron interaction mediated by the tunneling of a spin-1 magnon. This term can also be written as $-J \mathcal{P}_{i,i+1}^{S=0}$, with $\mathcal{P}_{i,i+1}^{S=0}$ being the projector onto the total singlet state of two neighboring electrons at sites $i$ and $i+1$. The third term ($V$) is a nearest neighbor repulsion, which can be interpreted as tunneling of a photon, and finally the last term ($U$) is the local charging energy.

\begin{figure}
\includegraphics[width=0.8\columnwidth]{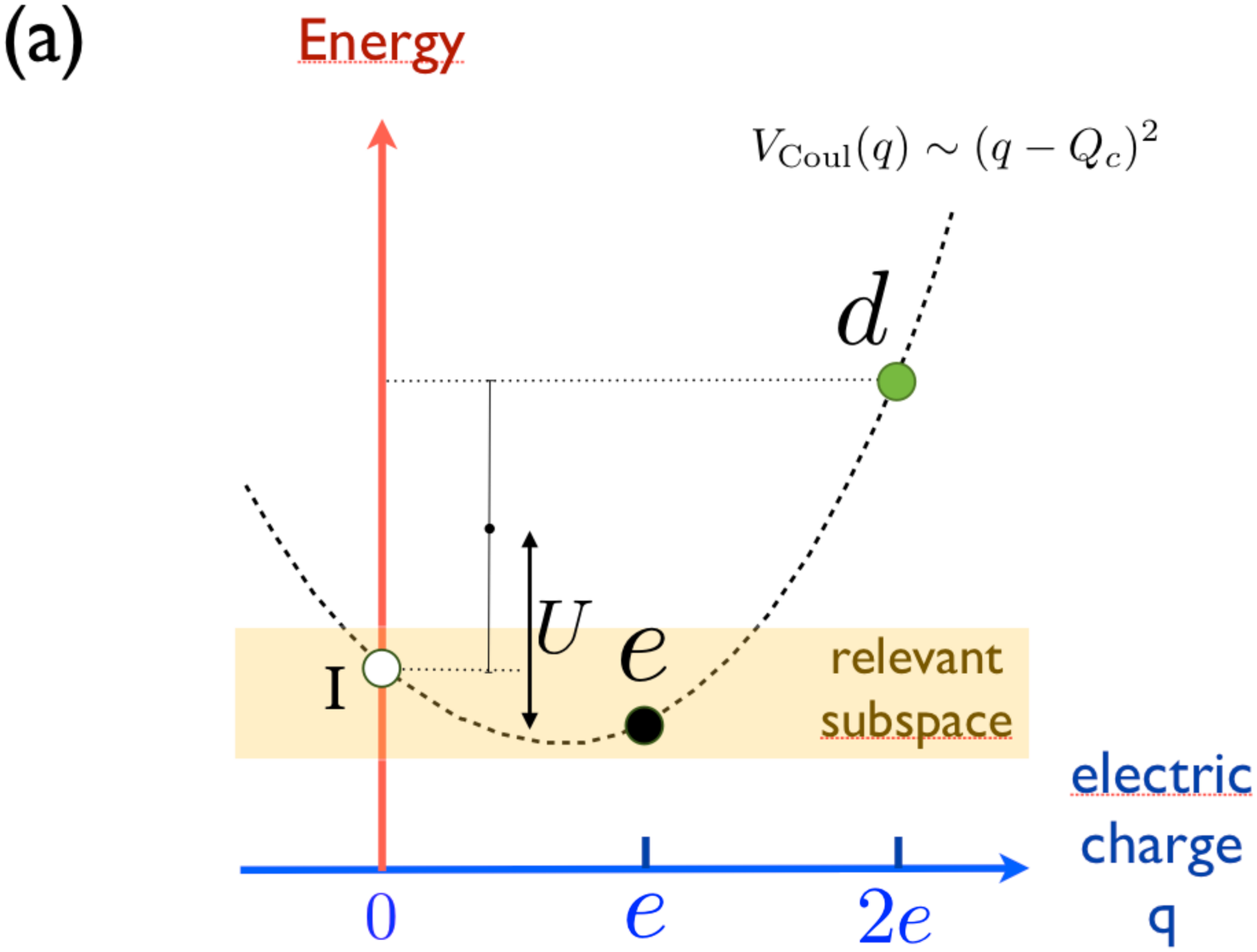}
\vskip 0.3truecm
\includegraphics[width=0.8\columnwidth]{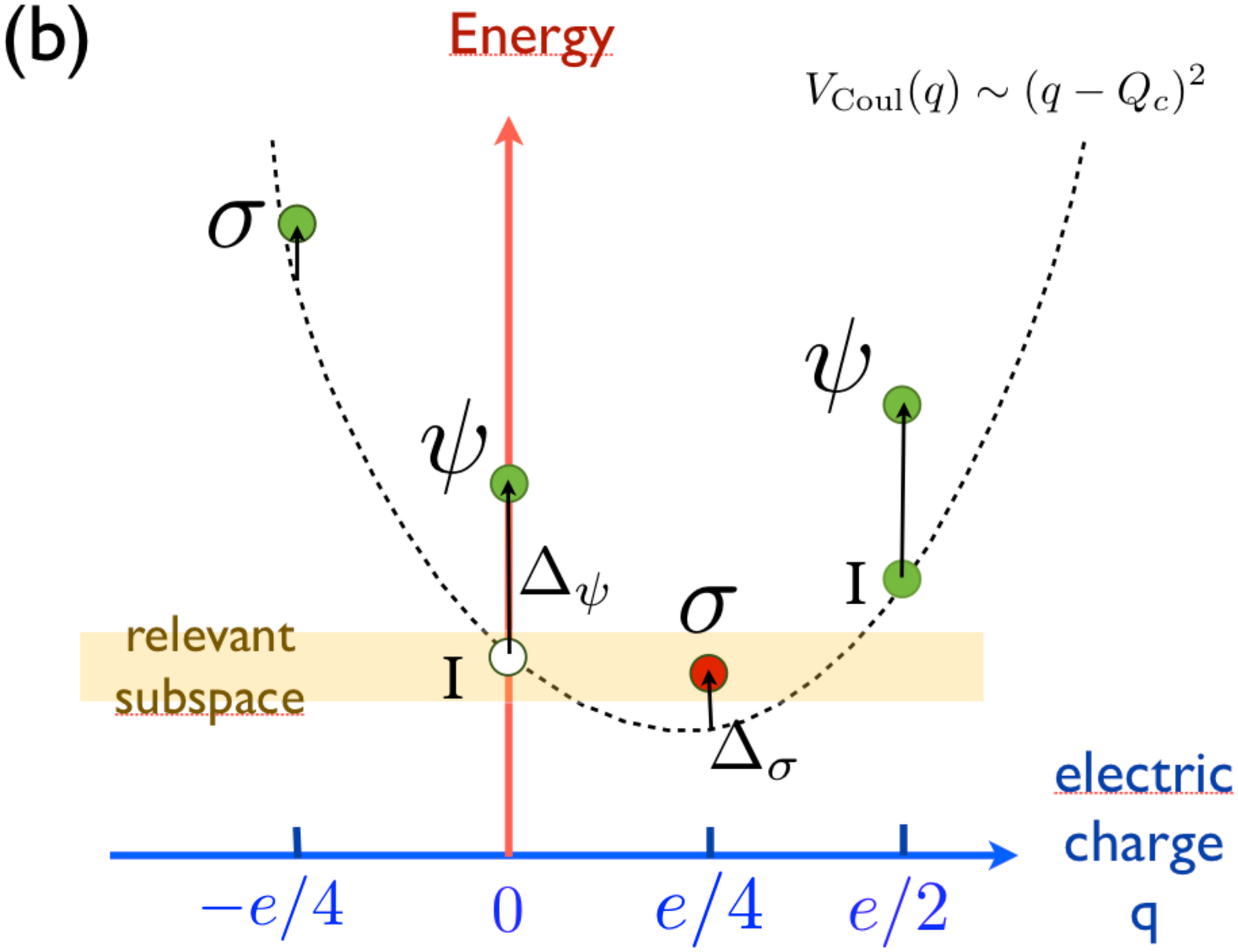}
\vskip 0.3truecm
\includegraphics[width=0.8\columnwidth]{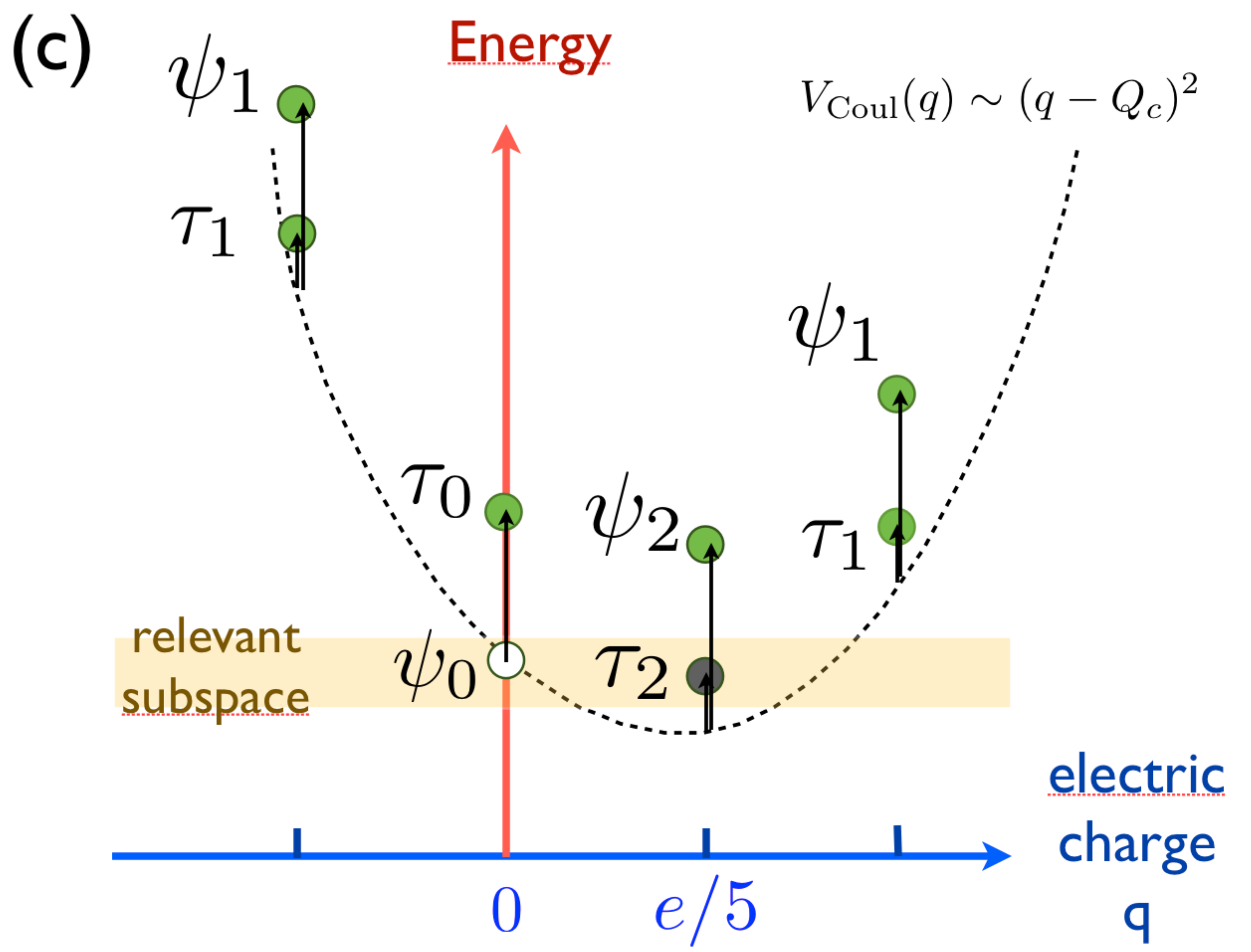}
\caption{Schematic energy spectra (arbitrary scale)
in the presence of a parabolic Coulomb charging energy
for the case of (a) electrons, (b) Ising anyons, and (c) Fibonacci anyons.
The Hubbard $U$ energy is shown in (a).
Upward arrows show the shifts corresponding to the topological contribution to the bare energies of the quasiparticles, e.g. $\Delta_\sigma=1/8$ and $\Delta_\psi=1$ in (b), $\Delta_{\tau_0}=4/5$,
$\Delta_{\tau_1}=\Delta_{\tau_2}=2/15$ and $\Delta_{\psi_1}=\Delta_{\psi_2}=4/3$ in (c). }
\label{fig:ener}
\end{figure}

\begin{figure}
\includegraphics[width=0.9\columnwidth]{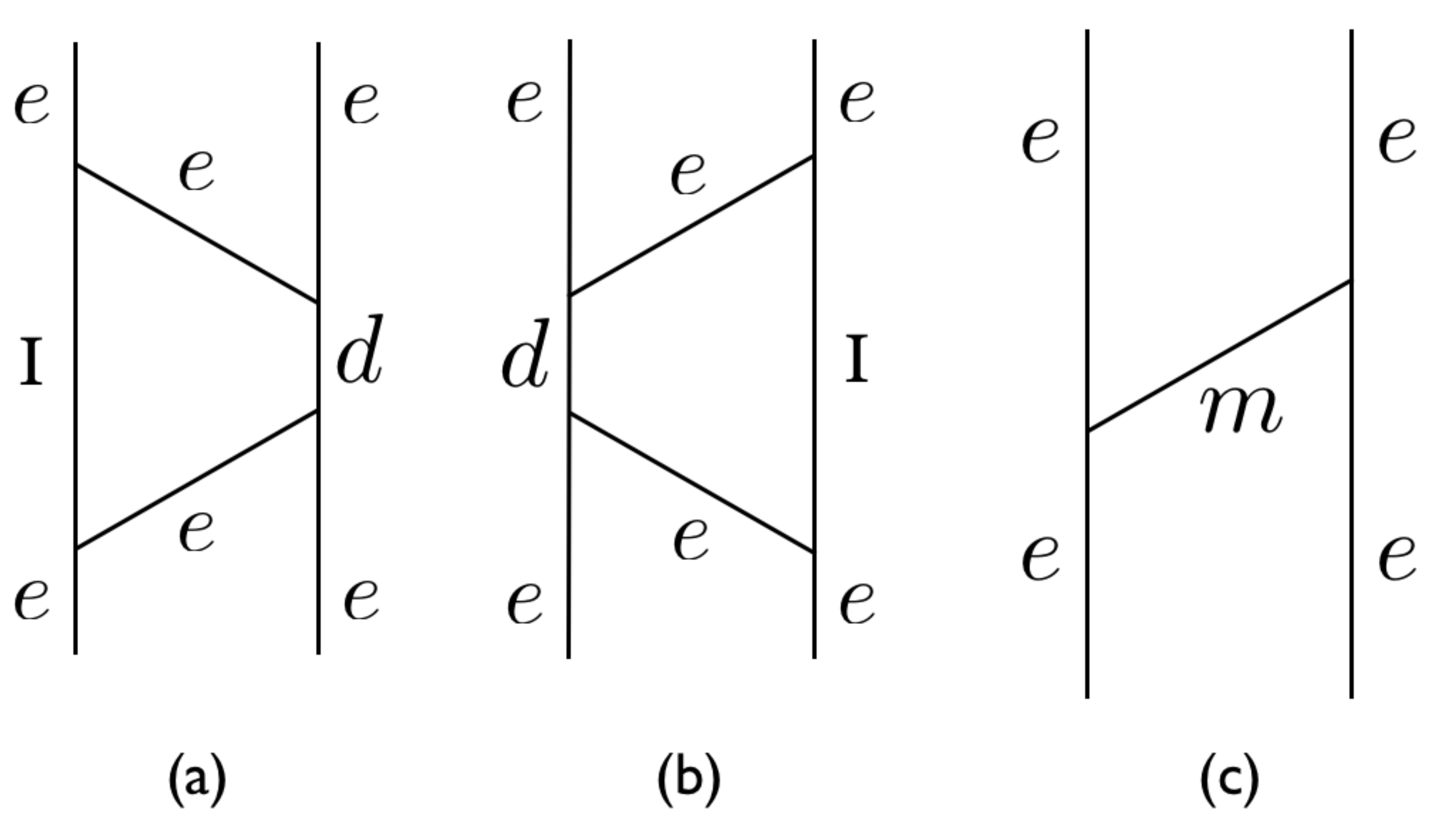}
\caption{(a,b) Second-order exchange processes for two electrons on nearest neighbor sites
via the virtual creation of a ``vacuum'' quasiparticle $I$ (i.e. a vacancy) and a ``doublon'' $d$.
These two exchange diagrams can be replaced by a first-order magnon exchange diagram, shown in (c), leading to a renormalization of the magnon exchange interaction.}
\label{fig:heis}
\end{figure}

In this Hubbard model of electrons, we consider three different types of ``quasiparticles'' at the lowest energies:
the ``trivial particle'' $I$ (i.e. a vacant site), the electron $e$, and the ``doublon'' $d$, corresponding to
a double (electronic) occupancy on a site. In Fig.~\ref{fig:ener}(a), we model the energy costs for a given quasiparticle type using a parabolic Coulombic charging energy of the form
\begin{equation}
V_{\text{Coul}}(q) \sim \left( q - Q_c \right)^2
\end{equation}
where $q$ is the quasiparticle's electric charge value, and $Q_c$ is the minimal energy charge value.
Here, each lattice site is viewed as a ``quantum dot" for which $Q_c$ is fixed by
the (implicit) chemical potential.
The last term in the Hamiltonian of Eq.~(\ref{eq:H_t_J}) specifies that $U$ is the energy cost for promoting two electrons into a vacancy and a doublon. When $U/t$ is large, one can project out doublons
and consider a restricted subspace of electrons and vacancies only.
The local $U$ interaction can then be taken into account in second-order perturbation, as shown in Fig.~\ref{fig:heis}, renormalizing the coupling constant of the spin exchange term to  $J=J_0+4t^2/U$, i.e. the magnon mediated interaction.

\subsection{Hilbert space truncation for anyons}

For anyons we proceed in the same way as for the electronic Hubbard model to derive a simpler effective model of the low lying states,
assuming that the charging energy is the largest energy scale and can be integrated out.
In Figs.~\ref{fig:ener}(b,c), we model the energy costs for a given quasiparticle type using a quantum dot, which again has a quadratic Coulombic charging energy $V_{\text{Coul}}(q)$, but also has an energy shift
\begin{equation}
V_{\text{neut}}(a) \propto \Delta_a
,
\end{equation}
where $\Delta_a$ is the conformal scaling dimension corresponding to topological charge $a$, which depends on the (neutral) topological charge of the quasiparticle~\cite{Bonderson10b}.
We note that conformal scaling dimension $\Delta_a$ is the sum of the left and
right conformal weights $\Delta_a=h_a+\bar{h}_a$.

For large charging energy we can restrict ourselves
to a low-energy subspace that contains only two quasiparticle types, as indicated in Figs.~\ref{fig:ener}(b,c).
In the case of the Ising anyon chain, we only allow the lowest energy quasiparticles $(I,0)$ and $(\sigma,e/4)$
to be localized on a given site. The quasiparticles $(I,e/2)$ and $(\psi,e/2)$ correspond physically to double
occupancies of the quasiparticle $(\sigma,e/4)$, and thus involve a prohibitively large Hubbard-like charging energy.
The neutral fermion quasiparticle $(\psi,0)$ is gapped, because of the energy associated with the $\psi$ mode, but it will be present in our model in the form of virtual tunneling processes.
To make the quasiparticles $(I,0)$ and $(\sigma,e/4)$ nearly degenerate, one has to introduce
a chemical potential, which combined with the charging energy gives the sought after near-degeneracy.
Similarly, in the case of Fibonacci anyons, we can also allow only the lowest energy quasiparticles $(\psi_0,0)$ and $(\tau_2,e/5)$ to be localized on a given site. In this case, the neutral Fibonacci quasiparticle $(\tau_0,0)$ is gapped, but will be present in our model in the form of virtual tunneling processes.
In Appendix~\ref{app:truncation}, we explain in more detail how one can combine the effects of a gate and of the charging energy in order to obtain a low-energy sector containing two degenerate states, separated from the other excitations by a gap.

We note that the fundamental quasiparticles of the MR or RR
quantum Hall states are described as a product of an Ising or Fib anyon model with an Abelian anyon model (see Appendix~\ref{app:anyonmodels})
which can be associated to the electric charge and is, therefore, ``additive.''
Hence, the electric charges of the quasiparticles of the relevant subspace
need not to be specified any more, because each quasiparticle has the same charge.
Also, we can drop the subscripts for the two lowest-energy Fibonacci quasiparticles, i.e. we identify
$\psi_0\rightarrow I$ and $\tau_2\rightarrow\tau$ since only one species
of $\tau$-anyons is involved in the low-energy subspace.
In other words, we end up with only one type of $\sigma$-anyon or $\tau$-anyon allowed on the sites
and the sites left empty are filled with trivial quasiparticles $I$ or vacancies.
The details of the above mentioned identification are not important here, but will be given in
Appendix~\ref{app:truncation}.

A pictorial representation of such chains was shown in Fig.~\ref{fig:sketch}.
The charge degrees of freedom can therefore be thought of as living on the sites (the localized nontrivial quasiparticles carry elementary
$e/4$ or $e/5$ electric charge, in contrast to the vacancies) while the bond variables $x_i$ are encoding the
anyonic (or spin) degrees of freedom. We recall that $a \times I = a$ for any quasiparticle
$a = I,\tau,\sigma,\psi$ so that the anyonic ``spin" is conserved along the empty segments of the chain (i.e. with vacancies on the sites).

\subsection{Interaction between nearest-neighbor anyons}

\begin{figure}
\includegraphics[width=0.9\columnwidth]{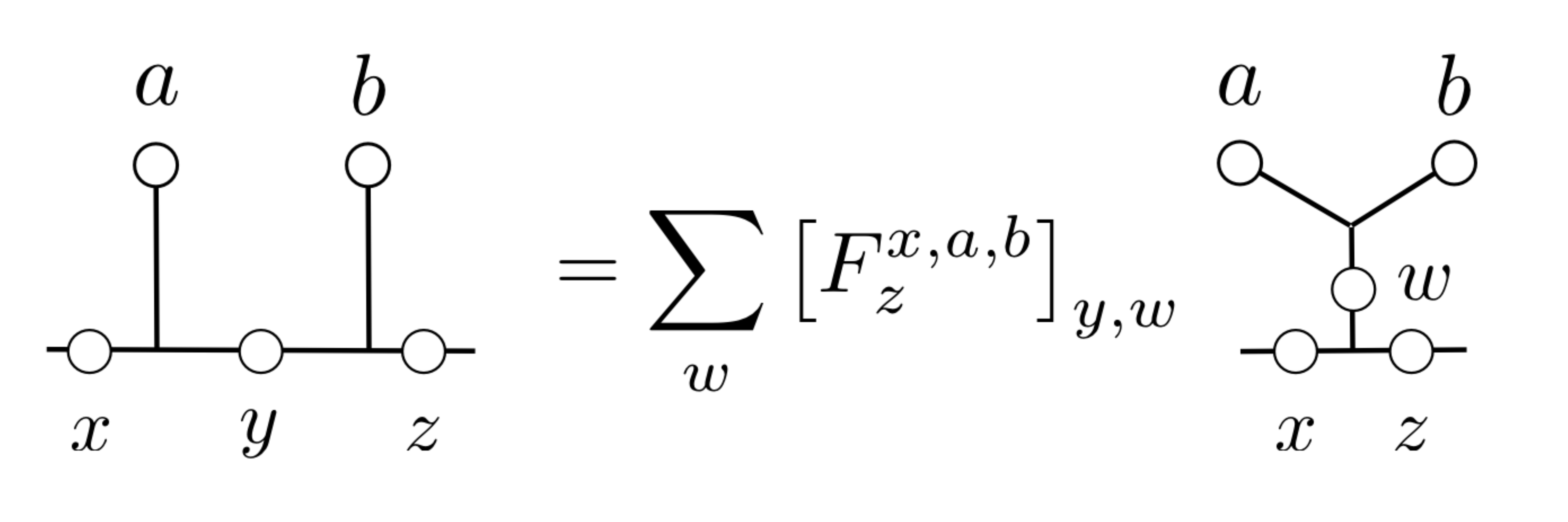}
\caption{Change of basis involving the fusion channel of two neighboring electrons or
anyons and the $F$-symbol. }
\label{fig:hJ}
\end{figure}

Let us consider putting $N$ anyons of type $\sigma$ or $\tau$ on an $L$-site chain
with periodic boundary conditions (PBC), i.e. on a closed ring (situated on a torus).
When two charged anyons sit on nearest-neighbor (NN) sites they experience
an ordinary Coulomb repulsion $V$. In addition, they interact via an effective exchange interaction of magnitude $J$,
which can be derived as in the electronic Hubbard chain.
For this, we use the (unitary) $F$-symbol transformation shown in Fig.~\ref{fig:hJ}, which is a change of basis between different fusion tree representations of the states. When we apply the $F$-symbol of the NN anyons, which have charges $a$ and $b$, it provides a change of basis from the fusion chain basis of Fig.~\ref{fig:sketch} (which we use to encode states) to one in which the fusion channel of this NN pair of anyons is manifest, as indicated in Fig.~\ref{fig:hJ} by the charge label $w$.

\begin{figure}
\includegraphics[width=0.9\columnwidth]{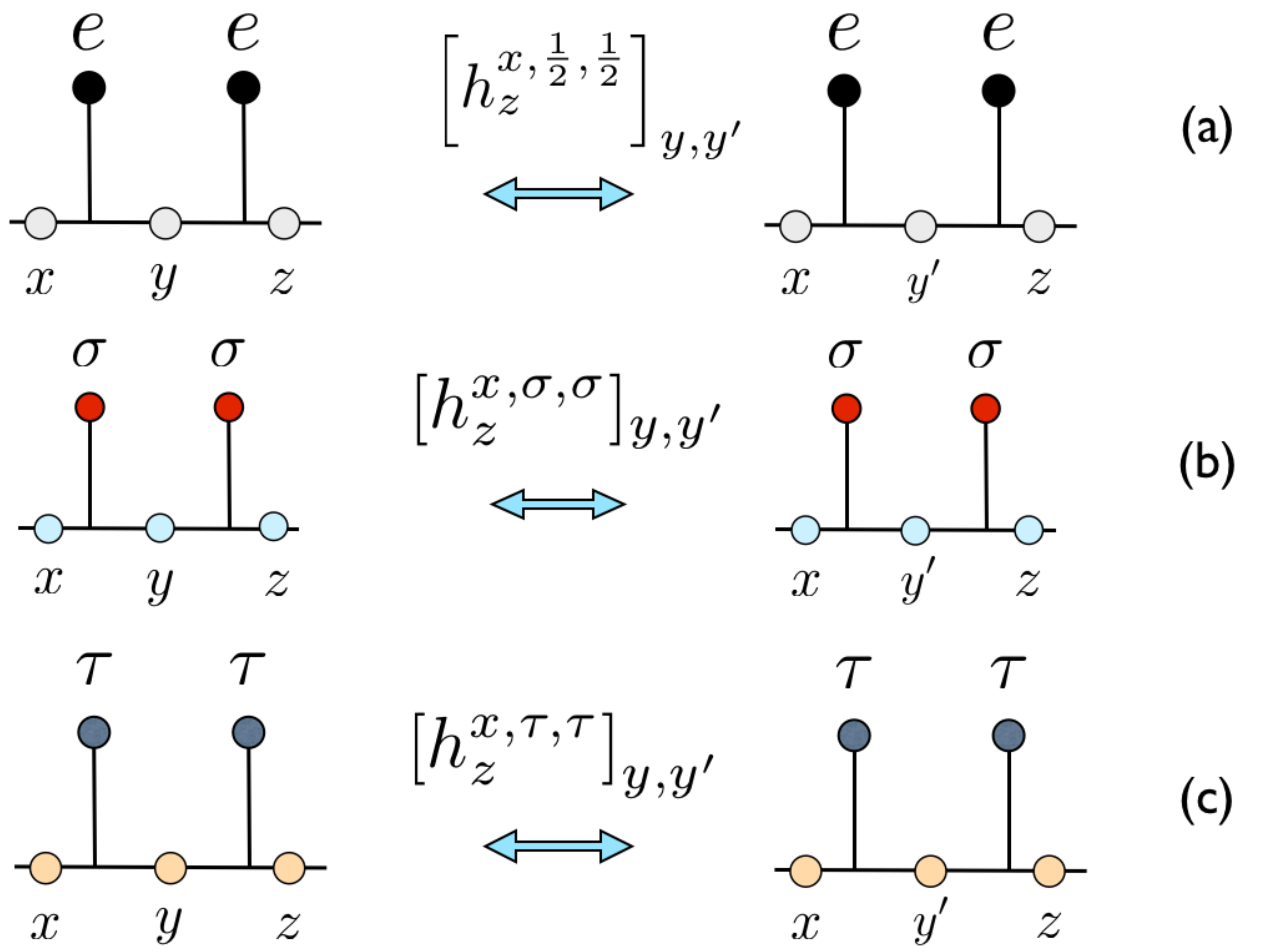}
\caption{Matrix elements describing (exchange) interactions between nearest neighbors of (a)
electrons, (b) Ising anyons, and (c) Fibonacci anyons.
In (b),  when $x=z=\sigma$, $y$ and $y'$ can take the values $I$ or $\psi$, making
$h^{\sigma,\sigma,\sigma}_\sigma$ a $2\times 2$ matrix, while when neither $x$ nor $z$ equals $\sigma$, then $y=y'=\sigma$, making $h^{x,\sigma,\sigma}_z$ a $1\times 1$ matrix.}
\label{fig:hJ2}
\end{figure}

By analogy with the electronic Heisenberg interaction,
the exchange interaction between two neighboring anyons is given by $-J{\cal P}^{I}$, which favors
the vacuum fusion channel $I$ for these two anyons. The action of the corresponding exchange processes on the local fusion tree basis elements are shown schematically in Figs.~\ref{fig:hJ2}(a-c).
By using the $F$-symbol change of basis of Fig.~\ref{fig:hJ}, the
local Hamiltonian elements $h^{x_{i-1},\alpha_{i},\alpha_{i+1}}_{x_{i+1}}$ can be derived, depending on (and labelled by) the
variables $x_{i-1}$ and $x_{i+1}$ on the two outer bonds connected to the two NN sites and
acting upon the local spin $x_i$ of the inner bond, as shown in Fig.~\ref{fig:hJ2}(a-c). The label $\alpha_{i}$
denotes the type of anyon localized at site $i$. Severe local constraints greatly reduce the number of possible non-zero
matrices and matrix elements, which we give explicitly below.

Let us first start with the case of two NN spin-1/2 (localized) electrons experiencing an AF exchange
interaction, i.e. for which the fusion outcome in the singlet channel is favored over the triplet channel.
In the usual spin-basis, this is just the Heisenberg term $- J (1/4 - {\bf S}_{i}\cdot {\bf S}_{i+1})$. However, we
work in the fusion chain basis, as pictured in Fig.~\ref{fig:sketch}(a). Thus, we need to know the
$F$-symbols describing the change of basis as given in Fig.~\ref{fig:hJ}, for the case of $SU(2)$ spin-$1/2$'s, i.e. $\alpha_{i} = \alpha_{i+1} = 1/2$.
The $F$-symbols are closely related to the Wigner $6j$-symbols [see, for instance Ref.~\onlinecite{book:messiah62} or the SU$(2)_k$ $F$-symbols with $q=1$ (i.e. $k=\infty$) in Appendix~\ref{app:anyonmodels}].
The $F$-symbols of interest here are given by
\begin{equation}
\begin{split}
&\left[ F^{x_{i-1},\frac{1}{2},\frac{1}{2}}_{x_{i+1}} \right]_{x_i,\tilde{x}_i} =
(-1)^{x_{i-1}+x_{i+1}+1}  \\
& \times \sqrt{(2x_{i}+1)(2\tilde{x}_i+1)}
\begin{Bmatrix}
x_{i-1} & 1/2 & x_{i} \\
1/2 & x_{i+1} & \tilde{x}_{i}
\end{Bmatrix}
\end{split}
\end{equation}
where $\tilde{x}_i=0,1$ is the total spin of the two spin-$\frac{1}{2}$'s, and
$\begin{Bmatrix} j_1 & j_2 & j_{12} \\ j_3 & j & j_{23} \end{Bmatrix}$
denote the $6j$-symbols.

In particular, if $x_{i+1} = x_{i-1} \pm 1$,
then the value of $x_{i}$ and $\tilde{x}_{i}$ are fixed to be $x_i = x_{i-1} \pm \frac{1}{2}$ and
$\tilde{x}_{i} = 1$, and the resulting $F$-symbol is just a number,
namely
\begin{equation}
\left[ F^{x_{i-1},\frac{1}{2},\frac{1}{2}}_{x_{i-1} \pm 1 } \right]_{x_{i-1} \pm \frac{1}{2}, 1} = 1
.
\end{equation}

In the case that $x_{i-1} = x_{i+1} = 0$, we must have $x_{i} = 1/2$ and
$\tilde{x}_i=0$. The associated $F$-symbol is again $1$,
\begin{equation}
\left[ F^{0,\frac{1}{2},\frac{1}{2}}_{0 } \right]_{\frac{1}{2}, 0} = 1
.
\end{equation}

The only case for which the $F$-symbol has rank two is when
$x_{i-1} = x_{i+1} = s \geq \frac{1}{2}$, giving $x_{i} = s \pm \frac{1}{2}$ and $\tilde{x}_{i} = 0,1$.
The $F$-symbol takes the from
\begin{equation}
F^{s,\frac{1}{2},\frac{1}{2}}_{s}  = \frac{1}{\sqrt{2s+1}}
\left[
\begin{matrix}
-\sqrt{s} & \sqrt{1+s} \\
\sqrt{1+s} & \sqrt{s}
\end{matrix}
\right]
\end{equation}
where the first column corresponds to $\tilde{x}=0$, and the second one $\tilde{x}=1$.

With the knowledge of the $F$-symbols, we can construct the Hamiltonian [see Fig.~\ref{fig:hJ2}(a)],
which symbolically takes the form
\begin{eqnarray}
\left[ h^{x_{i-1},\frac{1}{2},\frac{1}{2}}_{x_{i+1}} \right]_{x_{i},x'_{i}} &=&
V \delta_{x_{i},x'_{i}}\\
&-& J \left[ F^{x_{i-1},\frac{1}{2},\frac{1}{2}}_{x_{i+1}}\right]_{x_{i},0} \left[ (F^{x_{i-1},\frac{1}{2},\frac{1}{2}}_{x_{i+1}})^{-1}\right]_{0,x'_{i}}
 \nonumber\end{eqnarray}
where we included the coulomb interaction $V$, because the electrons occupy neighboring sites,
and we favor the spin-$0$ channel (implicitly it is only non-zero if the diagram is allowed by the fusion rules).
Explicitly, we find that $h^{x_{i-1},1/2,1/2,}_{x_{i+1}} = V$ in the case that $x_{i+1} \neq x_{i-1}$.
For $x_{i-1}=x_{i+1}=0$, we have $h^{0,1/2,1/2}_{0} = V - J$, and for $s > 0$, we have
\begin{equation}
h^{s,1/2,1/2}_{s} =
\left[
\begin{matrix}
V-\frac{J s}{2s+1} & \frac{J\sqrt{s(1+s)}}{2s+1} \\
\frac{J\sqrt{s(1+s)}}{2s+1} & V-\frac{J (s+1)}{2s+1}
\end{matrix}
\right] \ .
\end{equation}

The Hamiltonian in the case of Ising and Fibonacci anyons [see Figs.~\ref{fig:hJ2}(b,c)] is obtained in the same way as we did above for
spin-$1/2$ electrons. The most important necessary ingredient are the $F$-symbols, which can
be found in Appendix~\ref{app:anyonmodels}.

For Ising anyons,
the non-zero matrix elements are limited to
\begin{eqnarray}
\left[h^{\id,\sigma,\sigma}_{\id}\right]_{\sigma,\sigma} &=&
\left[h^{\psi,\sigma,\sigma}_{\psi}\right]_{\sigma,\sigma} = V-J \\
\left[h^{\id,\sigma,\sigma}_{\psi}\right]_{\sigma,\sigma} &=&
\left[h^{\psi,\sigma,\sigma}_{\id}\right]_{\sigma,\sigma} = V
\end{eqnarray}
and
\begin{equation}
h^{\sigma,\sigma,\sigma}_{\sigma} =
\left[
\begin{matrix}
V-J/2 & -J/2\\
-J/2 & V-J/2
\end{matrix} \right] \, ,
\end{equation}
where the basis used to write the matrix is $\{\id,\psi\}$.

The non-zero matrix elements of the Fibonacci chain are given by
\begin{eqnarray}
\left[ h^{\id,\tau,\tau}_{\id} \right]_{\tau,\tau} &=& V-J \\
\left[ h^{\id,\tau,\tau}_{\tau} \right]_{\tau,\tau} &=&
\left[ h^{\tau,\tau,\tau}_{\id} \right]_{\tau,\tau} = V
\end{eqnarray}
and
\begin{equation}
h^{\tau,\tau,\tau}_{\tau} =
\left[
\begin{matrix}
V-J/\phi^2 & -J/\phi^{3/2}\\
-J/\phi^{3/2} & V-J/\phi
\end{matrix} \right] \, ,
\end{equation}
where $\phi$ is the golden ratio and the matrix is written in the basis $\{I,\tau\}$.

\subsection{Anyon ``hopping"}

\begin{figure}
\includegraphics[width=0.9\columnwidth]{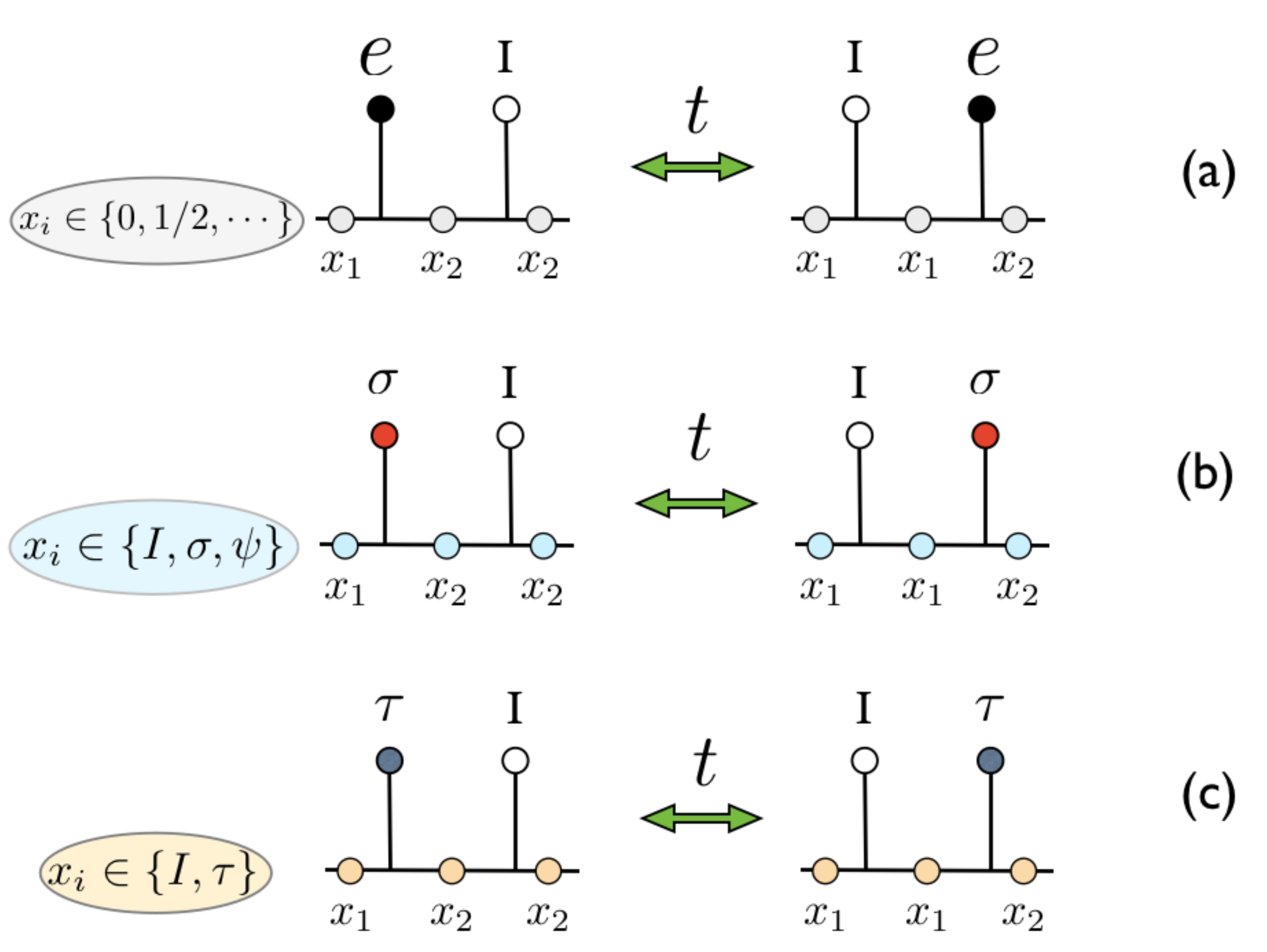}
\caption{ Tunneling process (or ``hopping") of (a) an electron, (b) an Ising anyon, and (c) a Fibonacci anyon. }
\label{fig:ht}
\end{figure}

Finally, we have to consider the possibility for quasiparticles (including the vacuum $I$) to move on the lattice and gain kinetic energy.
The (effective) physical hopping processes are shown in Fig.~\ref{fig:ht}.
In such a move, the entire quasiparticle, including the electric charge and spin/topological charge, is transported from one lattice site to a vacant site that is adjacent to it. (Generally, ``hopping'' may involve transfer of a quasiparticle to an occupied site, but we do not consider such processes in our models.)
Note that the magnitude of the hopping $t$ is not affected by the truncation
of the Hilbert space to the reduced space of the low-energy quasiparticles.
Note also that the sign of $t$ is irrelevant, so one can assume $t>0$ for simplicity.

When $|J|$ is large in comparison to $t$ and $V$, the system phase-separates i.e. the anyons
tend to form large clusters of higher density $\rho\simeq 1$. In contrast, for larger kinetic energy (i.e. $t$) and/or repulsion between the anyons, the system remains homogeneous.
This is the regime of interest
here and, for $\rho=2/3$ and $\rho=1/2$, we have found that it is realized for $t=|J|$, even when $V=0$,
or larger $t/|J|$ values.
Note that for convenience we assume $V=0$ throughout
and we have explored the models for values of $J$ ranging from $-t$ to $t$.

\section{Charge degrees of freedom for $J=0$}

This section is a ``warm-up" for the real itinerant anyon models,
starting with the simple example of identical bosons, and describes how making them
distinguishable introduces a twist in the periodic boundary conditions.
It therefore explains the $J=0$ part of the spectra, without the non-Abelian complications.

\subsection{Hard-core bosons}

\begin{figure}
\includegraphics[width=0.9\columnwidth]{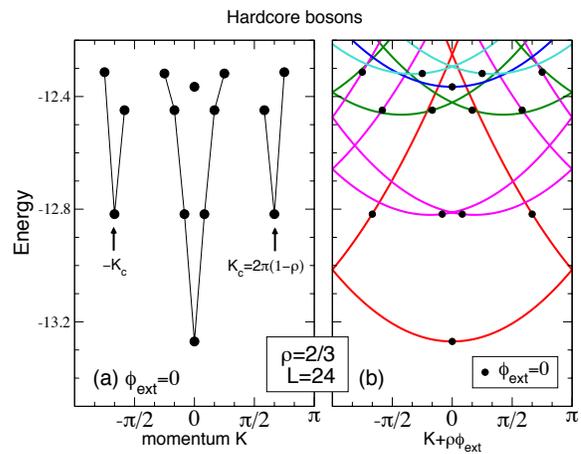}
\caption{Spectra of 16 (hardcore) bosons moving on a 24-site chain with PBC. (a) Spectrum
for zero magnetic flux through the ring. The linear dispersions vs momentum $K$
are shown at $K=0$ and $K=\pm K_c$. (b) Spectrum vs. external (continuous) magnetic flux
$\phi_{\rm ext}$ (multiplied by the density $\rho$). Each
discrete level (black dots) leads to a (parabolic) branch of excitations.}
\label{fig:HCB}
\end{figure}

We start with a periodic chain of size $L$ filled with $N$ bosons.
We consider the case where two bosons cannot occupy the same site
due e.g. to an infinite on-site repulsion (hard-core constraint).
Such a system of hard-core bosons (HCBs) can be mapped via
a Holstein-Primakoff and Jordan-Wigner  transformation onto a gas of
spinless fermions. In 1D, the effect on the spectrum due to the
difference in statistics can simply be accounted for by adding to the fermions an
extra phase shift of $\pi$ through the ring
(when the particle number $N=\rho L$ is even).
Therefore, the HCB many-body spectrum can be obtained by filling up $N$ states of a (fermionic) cosine band,
\begin{equation}
E_{\rm HCB} (p) = -2t\sum_{j(p)}\cos \left[\frac{2\pi}{L}(j+\frac{1}{2})\right] \, ,
\label{Eq:HCB-1}
\end{equation}
where $\{j(p)\}$ is a set (labelled by $p$) of an even number $N$ of different integers
and the momenta are all shifted by $\pi/L$.
The spectrum (for $t=1$) is displayed in Fig.~\ref{fig:HCB}(a).
As expected, the HCB spectra exhibit linear quasiparticle dispersions
centered at momenta $K=0$ and $K=K_c\equiv 2\pi \rho$ [or $2\pi (1-\rho)$ for $\rho> 1/2$].

For later use in the case of anyons, it is of interest to introduce an external magnetic
flux $\phi_{\rm ext}$  or equivalently an Abelian U(1) flux through the ring.
The new spectrum $E_{\rm HCB}(p,\phi_{\rm ext})$
depends now on both discrete and continuous variables $p$ and $\phi_{\rm ext}$,
\begin{equation}
E_{\rm HCB}(p,\phi_{\rm ext}) = -2t\sum_{j(p)}\cos\left[ \frac{2\pi}{L}(j+\frac{1}{2})+\frac{\phi_{\rm ext}}{L} \right] \, ,
\label{Eq:HCB-2}
\end{equation}
plotted in Fig.~\ref{fig:HCB}(b) (for $t=1$).The state labelled by $(p,\phi_{\rm ext})$ now carries an arbitrary (continuous)
total momentum ${\tilde K}=K_p+\rho\phi_{\rm ext}$ with $K_p=\frac{2\pi}{L}\sum_{j(p)}(j+\frac{1}{2})$.

\subsection{Fermions and anyons}

We now make the (quasi)particles {\it distinguishable}, i.e. we introduce some internal degrees of freedom
which can be e.g. the spin-1/2 components of the electrons or the anyonic degrees of freedom of
the Ising $\sigma$ or Fibonacci $\tau$ anyons. The resulting model is the same as considering the
above $t$-$J$ models in the limit where $J=0$ ($t$ can be set to 1).
This limit, where the energy scale of the anyonic degrees of freedom
are set to zero, is of great interest since it provides insight on the nature of the charge excitations.

\begin{figure}
\vskip 0.3truecm
\includegraphics[width=0.9\columnwidth]{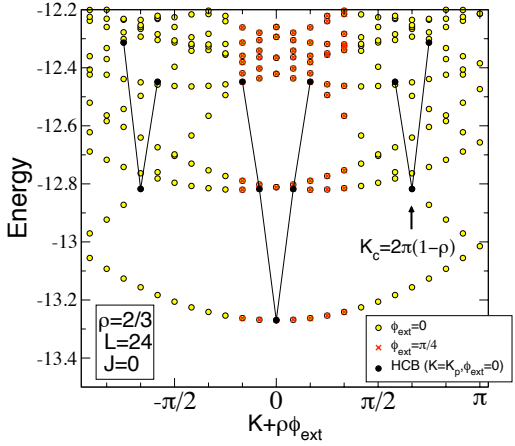}
\caption{Low-energy spectrum of a 24-site $t$-$J$ chain at $J=0$ and density $\rho=2/3$, obtained numerically using Ising anyons.
Yellow circles correspond to the spectrum at $\phi_{\rm ext}=0$ as a function of
momentum $K$. The ``parent" charge excitations (HCB at $\phi_{\rm ext}=0$) are
shown by full (black) circles. Adding an external flux $\phi_{\rm ext}$ through the ring is
equivalent to shift $K$ by $\rho\phi_{\rm ext}$~:
the (red) $\times$ symbols correspond to the spectrum at $\phi_{\rm ext}=\pi/4$ restricting
$K\in[-\pi/3,\pi/6]$ (so that ${\tilde K}\in[-\pi/6,\pi/3]$).
}
\label{fig:Maj0}
\end{figure}

In that limit, the mobile anyons are expected to still behave as HCB.
However, the extra internal degrees of freedom (with zero energy scale)
should provide extra features on top of the HCB spectrum.
Because of the anyonic (or spin) degrees of freedom linked to them
the charged bosons are no longer indistinguishable particles and, on a torus or a ring,
hopping of a particle across the ``boundary'' cyclically translates the labels of the fusion tree.
To recover the same labeling, in general, all $N$ particles must
be translated over the boundary. Thus, one distinguishable particle
hopping over the boundary has the same effect as a phase shift $\phi_n=2\pi \frac{n}{N}$ (with $n$
an integer).
Hence the complete $J=0$ electronic/anyonic spectrum (at zero external flux)
is given by the union of all HCB spectra taken at all discrete values of $\phi_n$,
\begin{equation}
E_{\rm charge}^{\, p,n}=E_{\rm HCB}(p,\phi_n) \, .
\end{equation}
A momentum shift is induced by the U(1) flux,
given by $\rho\phi_n$ i.e.
$2\pi \frac{n}{L}$, an integer multiple of $\frac{2\pi}{L}$. The states then
carry (discrete) total momenta,
\begin{equation}
K_{p,n}=K_p+2\pi \frac{n}{L} \, ,
\end{equation}
where $K_p$ are the momenta of the HCB eigenstates at $\phi_{\rm ext}=0$.
For convenience one can distribute the phase shift
equally on the bonds to preserve translational invariance and one
gets,
\begin{equation}
E_{\rm charge}^{\, p,n}=-2t\sum_{j(p)}\cos\left[ \frac{2\pi}{L}(j+\frac{1}{2}+\frac{n}{N}) \right] \, .
\label{Eq:charge0}
\end{equation}

\begin{figure}
\vskip 0.3truecm
\includegraphics[width=0.9\columnwidth]{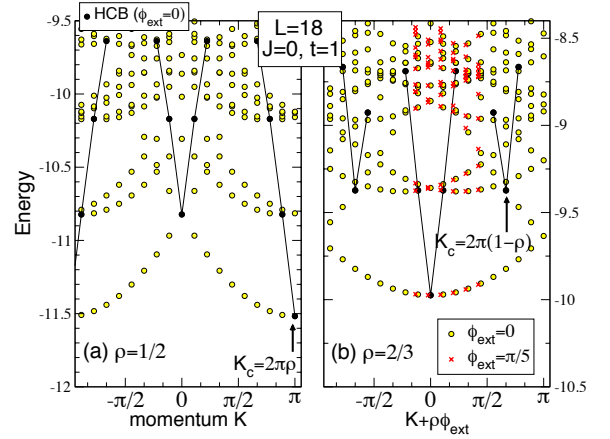}
\caption{Low-energy spectra vs. momentum $K$ of a 18-site $t$-$J$ chain with $J=0$ at
anyon densities (a) $\rho=1/2$ and (b) $\rho=2/3$. The notations here are the same
as in Fig.~\protect\ref{fig:Maj0}, but these results are obtained numerically using Fibonacci anyons.
The low-energy spectra of the 18-site HCB chain at the same densities (parent excitations)
are shown by black dots. Data for an external flux $\phi_{\rm ext}=\pi/5$ are shown in (b).
The minimum of the spectrum occurs at
momentum $K=0$ or $K=\pi$, depending on the parity of the number $N=\rho L$ of quasiparticles.
}
\label{fig:Fib0}
\end{figure}

Spectra for Ising and Fibonacci anyons obtained by exact diagonalization
(ED) for $J=0$ are shown in Fig.~\ref{fig:Maj0} and Fig.~\ref{fig:Fib0}(a,b).
As expected, Eq.~(\ref{Eq:charge0}) matches exactly the numerical exact diagonalization  results.
From the above considerations, it is then clear that the eigen-energies
lie exactly on top of the parabolas corresponding to the ``optical'' excitations
of the HCB. In other words, each state in the HCB spectrum is
extended into a discrete set of levels on a parabola -- the same parabola that
one gets by adding flux (an Abelian phase), as checked numerically.

It is important to notice that the $J=0$ spectrum does not depend on the internal degrees of freedom
and, hence, on the nature of the quasiparticles,
i.e. whether they are electrons, Ising anyons, Fibonacci anyons, or distinguishable bosons.
However, the respective Hilbert spaces are very different
which means that the corresponding eigenfunctions and degeneracies differ completely.
In addition, the way the very large degeneracy of each level is lifted by
any finite exchange interaction (see Fig.~\ref{fig:Maj0_smallJ} discussed later) depends crucially on the type of particles.

\subsection{External magnetic flux}

When the anyons experience an arbitrary {\it external} flux $\phi_{\rm ext}$, the above formula can be generalized to
$E_{\rm charge}^{\, p,n}(\phi_{\rm ext})=E_{\rm HCB}(p,\phi_n+\phi_{\rm ext})$.
It then becomes apparent that the $J=0$ energy spectrum does not depend
on the momentum $K_{p,n}$ and external flux $\phi_{\rm ext}$ separately but rather only on
the ``pseudo-momentum" combination ${\tilde K}=K_{p,n}+\rho\phi_{\rm ext}$. Hence, one can define a spectrum
depending on both discrete and continuous variables,
\begin{equation}
E_{\rm charge}(p,\Phi)= E_{\rm HCB}(p,\Phi)\, .
\label{Eq:charge1}
\end{equation}
The curvature of the ground-state energy $\partial^2 E_{\rm charge}(0,\Phi)/\partial\Phi^2$
is directly proportional to the optical (Drude) weight quantifying the potential of this
system to conduct.

\section{Dense anyon models}

To complete the warmup to describe the full anyonic $t$-$J$ model, we briefly discuss the
dense anyon models at $\rho = 1$. These models have precisely one anyon per site, which are, hence, immobile due to the
hard-core constraint. Every anyon interacts with its two neighbors, and only
the sign of the interaction strength $J$ is relevant. These models, introduced in Ref.~\onlinecite{ftl07},
are the anyonic versions of the Heisenberg spin chains. We will only consider the spin-$1/2$
versions in this paper.

In the case where only a nearest-neighbor two-body interaction is present, the spin-$1/2$ anyonic
chains are all critical, and their energy spectra are described by well known conformal field theories.
Starting with the Ising anyons, we note that, due to the fusion rules, the degrees of freedom on the
fusion chain are forced to form a pattern of alternating frozen $\sigma$ bonds and bonds fluctuating
between $I$ and $\psi$. For these later bond variables, the interactions of Fig.~\ref{fig:hJ}(c)
are exactly those of a {\it critical Ising model in transverse field} whose corresponding CFT has
central charge $c=1/2$. This is irrespective of the overall sign of the interaction, although the momenta
at which the various states occur differ depending on the sign of $J$.

In the case of Fibonacci anyons, changing the sign of $J$ does alter the critical behavior of the
chain. In the case of anti-ferromagnetic interactions (favoring the trivial fusion channel of two
neighboring Fibonacci anyons), the critical behavior is described in terms of the $c=7/10$ tri-critical Ising
model, with low-lying, linearly dispersing modes occurring
at momenta $K=0$ and $K=\pi$. For ferromagnetic interactions, the critical behavior is instead
described by the $c=4/5$ 3-state Potts model, which exhibits low-lying modes at
$K = 0,2\pi/3,4\pi/3$.

\begin{figure}
\vskip 0.3truecm
\includegraphics[width=0.9\columnwidth]{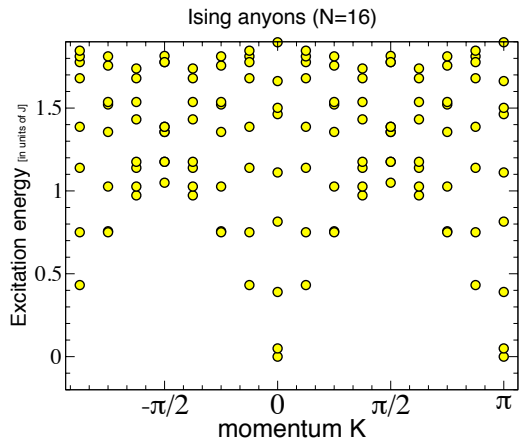}
\caption{Energy spectrum of a dense Ising-anyon chain ($\rho=1$) of length $L_a=16$.
The ground state energy has been subtracted.
}
\label{fig:majorana_L16}
\end{figure}

The behavior described above can be obtained by mapping the models onto exactly solvable
two-dimensional height models, introduced by Andrews, Baxter and Forrester.\cite{abf84} The
result for spin-$1/2$ anyons associated with SU$(2)_k$ is that in case of anti-ferromagnetic
interactions, the critical behavior of the chain is given in terms of the $k$-critical Ising model,
while ferromagnetic interactions give rise to the critical behavior of the $\mathbb{Z}_k$-parafermions.

We would like to stress that, although one can analytically obtain the critical behavior of the
dense anyonic chains, it is not possible to obtain the energy spectra in full detail for finite size
systems for $k\geq 3$. To obtain these, one must employ numerical techniques, such as exact diagonalization
(see e.g. Fig.~\ref{fig:majorana_L16}, for the $k=2$ Ising case, which can also be
obtained exactly). In describing the full spectra of the anyonic $t$-$J$ models, we will make
use of the spectra of the dense anyonic models described here, as obtained from exact
diagonalization.

We will denote the length of the dense anyon chains by $L_a$. The
energies of the dense anyon chains are denoted by $E_{\rm anyon} (m)$, where the integer $m$
labels the eigenstates, which have momenta $k_m$ that are integer multiples of $2\pi/L_a$.
Next, considering chains of length $L$ and at anyon densities $\rho< 1$,
we shall define the corresponding ``squeezed chains" of dense anyons of length $L_a = \rho L = N$, in which the
vacancies (or trivial quasiparticles) have been removed.

\section{Many-body spectra of the anyonic $t$-$J$ models}

Having described the spectra of the HCB system, in the presence of external flux, as well as the
spectra of dense anyon models, we are now ready to describe the spectra of the full, itinerant
anyon models. We will label the various energies as $E_{p,m}$, where the labels $p$ and $m$
refer to the (renormalized) HCB spectrum and the dense anyon chain, respectively. We will also explain the
subtle coupling of the momenta.

\subsection{Separation of charge and anyonic degrees of freedom}

We now consider to the full $J>0$ spectra of the itinerant
models. We solve the effective anyonic models on small periodic rings using exact diaginalization.
For Ising anyons, a 24-site chain is studied at density $\rho=2/3$ ($N=16$ quasiparticles).
For Fibonacci anyons (which have a larger Hilbert space) we consider a 18-site chain at density
$\rho=1/2$ ($N=9$ quasiparticles) and $\rho=2/3$ ($N=12$ quasiparticles).
We choose $|J|\leq t$ for which the system
remains uniform and does not phase separate (which occurs for larger $J$).
Note that the sign of $J$ is irrelevant for the energy levels in the case of Ising anyons,
though the momenta at which the various states occur differs depending on the
sign of $J$.

\begin{figure}
\vskip 0.3truecm
\includegraphics[width=0.9\columnwidth]{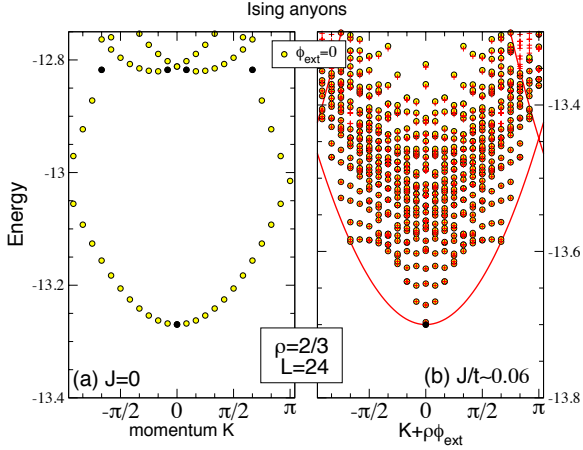}
\caption{A zoom-in on the low-energy spectra vs. $K$ of a 24-site Ising anyon $t$-$J$ chains at density $\rho=2/3$ for
(a) $J=0$ and (b) a small value of $J/t=\tan{(\pi/50)}\simeq 0.06$.
The Lanczos algorithm  with 800 iterations is used so that,
in the shown energy window, most of the eigen-energies have converged to within a relative error of $10^{-16}$
(a few not-fully converged levels are not shown).
The (red) $+$ symbols correspond to the sum of the (computed) lowest charge branch (continuous red line)
with {\it all} the expected anyonic excitations. (See text for more details.)
}
\label{fig:Maj0_smallJ}
\end{figure}

\begin{figure}
\vskip 0.3truecm
\includegraphics[width=0.9\columnwidth]{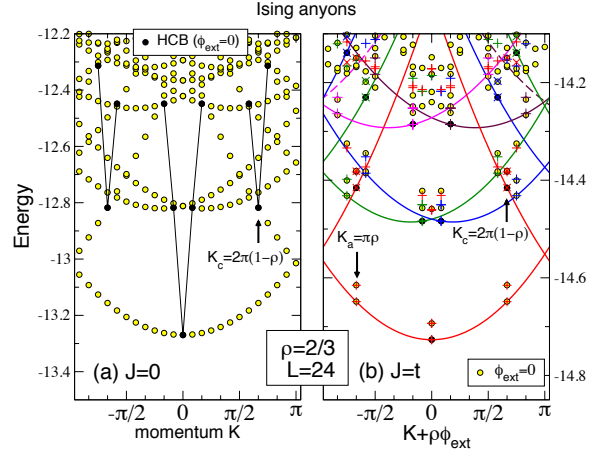}
\caption{Low-energy spectra vs. $K$ of a 24-site Ising $t$-$J$ chain at density $\rho=2/3$
for (a) $J=0$ and (b) $J=t=1/\sqrt{2}$. The data at $\phi_{\rm ext}=0$ are shown by yellow circles.
The parent charge excitations are shown by black dots ($\phi_{\rm ext}=0$) and, as a function
of pseudo-momentum ${\tilde K}=K+\phi_{\rm ext} \rho$ (varying $\phi_{\rm ext}$), by continuous lines
of different colors.
The $+$ and $\times$ symbols added for comparison correspond to the sum of the charge and
expected anyonic excitation spectra. (See text for more details.) The colors of these symbols
are the same as their {\it parent} charge excitation parabolas.
}
\label{fig:Maj}
\end{figure}

\begin{figure}
\vskip 0.3truecm
\includegraphics[width=0.9\columnwidth]{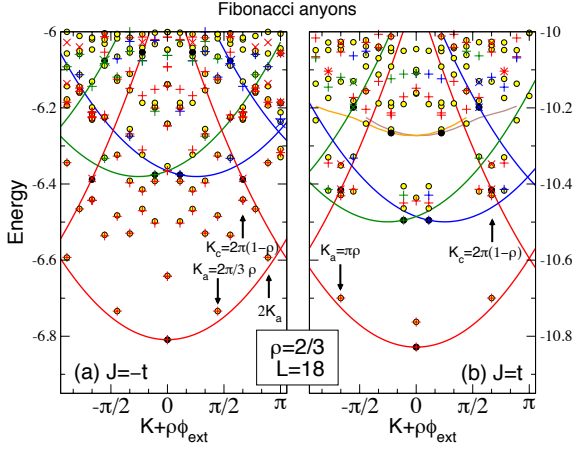}
\caption{Low-energy spectra vs. $K$ of a 18-site Fibonacci $t$-$J$ chain at density $\rho=2/3$
and $|J|=t=1/\sqrt{2}$ for both (a) $J<0$ and (b) $J>0$. The data at $\phi_{\rm ext}=0$ are shown by yellow circles.
The parent charge excitations are shown by black dots ($\phi_{\rm ext}=0$) and, as a function
of pseudo-momentum ${\tilde K}=K+ \rho\phi_{\rm ext}$ (varying $\phi_{\rm ext}$), by continuous lines
of different colors.
The $+$ and $\times$ symbols added for comparison correspond to the sum of the charge and
(expected) anyonic excitation spectra (see text). The color of these symbols
is the same as their {\it parent} charge excitation parabola.
}
\label{fig:Fib}
\end{figure}

The low-energy spectra ($|J|=t=1/\sqrt{2}$) of the itinerant Ising and Fibonacci anyonic chains
are shown in Figs.~\ref{fig:Maj}(b) and Fig.~\ref{fig:Fib}(a,b), respectively. These seem very different from the $J=0$ limit studied above
(and shown again in Fig.~\ref{fig:Maj}(a) for comparison).
To understand such spectra, let us first consider
a zoom-in on the low-energy region and compare the spectra at $J=0$ and at a small value of $J$,
as shown in Figs.~\ref{fig:Maj0_smallJ}(a,b).
This reveals that the {\it highly degenerate} $J=0$ charge excitation parabola is being split by
the magnetic interaction into a complex spectrum with a spread in energy proportional to $JL$.
When $J\sim t/L^2$, the spectra originating from each parabola start to overlap
as expected in Figs.~\ref{fig:Maj}(b) and \ref{fig:Fib}(a,b).
Despite the apparent complexity of the $J\ne 0$ spectrum, we shall be able to
express all excitations as the sum of an anyonic excitation and
a charge excitation, extending the concept of spin-charge separation familiar for 1D correlated electrons to the case
of a 1D anyonic interacting system. To complete this task, we first establish from simple considerations the ``recipes'' to
construct separately the expected charge and anyonic spectra. In a second step, we show how
the numerical spectra of the $t$-$J$ anyonic chains can be seen as a subtle combination of the
above two spectra.

The Bethe ansatz results~\cite{ogata-shiba,shiba-ogata,parola-sorella-92}
for the $J\rightarrow 0$ {\it electronic} $t$-$J$ chain suggest that
the anyonic contributions, $E_{\rm anyon} (m)$, to the excitation spectrum
of the itinerant anyon chain are those of the {\it squeezed} periodic chain of localized anyons produced by removing all vacant sites, which has the resulting length $L_a=N=\rho L$. Here, the integer $m$ labels the eigenstates of momenta $k_m$, which are multiples
of $2\pi /L_a$.
Such a spectrum can be computed separately by exact diagonalization
and agrees very well with the CFT predictions, even on small chains ($L_a=12, 16$).
In particular, it shows a (linear) zero energy mode at zero momentum and at a characteristic momentum
$k_a$, where $k_a=\pi$ for Ising and $J>0$ Fibonacci chains and $k_a=2\pi/3$ for $J<0$
Fibonacci chains.
The coupling constant providing the scale of the anyon spectrum is expected to be
weakly renormalized
from $J$ to $\gamma J$ in the doped system, where $\gamma$ is
a factor of order $1$ that is to be adjusted as we described below.

To construct the expected charge excitation spectrum at finite $J$,
we use our understanding of the charge excitations in the $J=0$ limit.
Starting from $J=0$ and turning on $J$ gradually,
one can {\it a priori} adiabatically follow the original ($\Phi=0$) HCB excitations evolving
in, what we call, the {\it parent} charge excitations at $J\ne 0$
(labelled by the same integers $p$ and at the same momenta $K_p$).
As for $J=0$, changing the momentum $K$ of a charge excitation amounts
to introducing a total phase shift (or flux) $\Phi=K/\rho$. Hence, by introducing ``twisted
boundary conditions" one can {\it compute} numerically
the (almost parabolic) branch of excitations ${\tilde E}_{\rm charge}(p,\Phi)$
associated to each parent excitation (labelled by $p$).
Note that  each branch is ``renormalized" by $J$ so that, strictly speaking, the charge spectrum is
no longer associated to {\it non-interacting} spinless
fermions (i.e. to HCBs) and, hence, is no longer given by
a simple analytic expression as in Eq.~(\ref{Eq:charge1}).
However, if different parent states lie on the same branch, they are still exactly spaced apart
by integer multiples of $\Delta\Phi=2\pi$ i.e.
${\tilde E}_{\rm charge}(p,\Phi)={\tilde E}_{\rm charge}(p',\Phi+2k\pi)$
where $K_{p'}-K_p= 2k\rho\pi$, for some integer $k$.

We now explain how to construct the full excitation spectrum by simply considering that
(i) the charge degrees of freedom
are subject to a phase shift in the boundary conditions and (ii) the anyonic degrees of freedom
are the ones of the {\it squeezed} periodic anyonic chain.
According to the above arguments, the
energy excitation spectrum should be given by adding the two contributions,
\begin{equation}
E_{p,m} = {\tilde E}_{\rm charge} (p, k_{m}) + E_{\rm anyon} (m) \, .
\label{eq:sp-ch-spec}
\end{equation}
A natural prescription is to simply add the momenta: $K=K_p+\rho k_m$.
In other words, we assume that the phase shift experienced by the charged ``holons" coincides with the total momentum $k_m=\frac{2\pi n_m}{L_a}$ (where $n_m$ is an integer) of the anyonic eigenstates defined on the squeezed (undoped) chain.
These rules for adding charge and anyonic momenta are therefore assumed to be similar
to the $J\rightarrow 0$ Bethe ansatz.

We now wish to verify that
proper assignments of the true energy levels according to the form
given by Eq.~(\ref{eq:sp-ch-spec}) can indeed be made accurately.
First, we consider adding a very small exchange coupling $J$ which
lifts the very large degeneracy of the low-energy parabola of the HCBs (see Fig.~\ref{fig:Maj0_smallJ}).
For finite $J$ the first charge branch (with $E_{\rm anyon}=0$) originating from
the zero-momentum ground-state of the model (which we assign $p=0$)
can be computed by adding an Abelian flux to the system.
It is then possible to construct the expected set of combined
charge plus anyon excitations $E_{0,m}$ by adjusting
the renormalization factor $\gamma$ to get the best fit to the exact low-energy levels.
Although there is only one free parameter, it is remarkable that {\it all} anyon excitations above the lowest
charge parabola can be assigned very accurately as seen in Fig.~\ref{fig:Maj0_smallJ}(b).

When $|J|\sim t$, charge and anyonic excitations have the same energy scale and one
must proceed step by step, sequentially constructing
the sets of levels corresponding to increasing charge index $p$.
The two ``secondary" parent charge excitations corresponding to exact eigenstates of the system
with momenta $K_p=\pm K_c$ ($p=1,2$) lie on
the {\it same} $p=0$ branch, as seen on Figs.~\ref{fig:Maj}(b) and \ref{fig:Fib}(a,b). These states
lead to the secondary level of combined excitations $E_{1,m}$ and $E_{2,m}$,
with no further adjustable parameter.
Recall that $K_c = 2\pi \rho$ if $\rho\leq 1/2$ and $K_c = 2\pi(1-\rho)$ for $\rho > 1/2$.
Next, in a second step, we identify the lowest {\it not yet assigned} excitations
at momenta $K_p=\pm 2\pi/L$
as the subsequent pure charge excitations (and assign them the labels $p=3,4$).
Following these levels adiabatically under the addition of
a flux enables us to construct the corresponding charge branches and locate the secondary
pure charge excitation at momenta $K_p=\pm (2\pi/L+K_c)$ (called $p=5,6$).
Then, as before, the combined excitations $E_{p,m}$, $p=3,...,6$, can be constructed.
One can repeat this procedure (going up in energy) until the level density
and the number of charge branch crossings becomes too large to make precise assignments.
In practice, we have identified up to $p=11$ pure charge excitations and
their corresponding low-energy combined anyonic-charge excitations
for the Ising chain, as shown in Fig.~\ref{fig:Maj}(b).
For the Fibonacci chains, we have identified up to $p=9$ pure charge excitations and
their corresponding low-energy combined anyonic-charge excitations, as shown in Fig.~\ref{fig:Fib}(a,b).

\begin{figure}
\vskip 0.3truecm
\includegraphics[width=0.9\columnwidth]{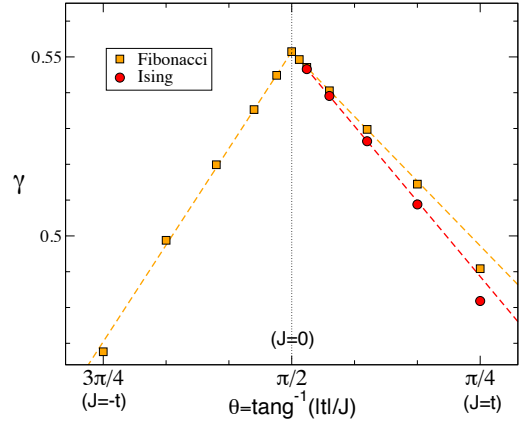}
\caption{Renormalization parameter $\gamma$ of the energy scale of the anyonic degrees of freedom
for Ising and Fibonacci at $\rho=2/3$, computed on $L=24$ and $L=18$ chains, respectively.
}
\label{fig:renormJ}
\end{figure}

Our results show that the anyonic energy spectrum is basically given by
the same type of Bethe ansatz as for the electronic $t$-$J$ model (in the $J/t\rightarrow 0$
limit)~\cite{ogata-shiba,shiba-ogata} and, in particular,
(i) the $J=0$ charge excitation spectrum {\it is exactly the same},
(ii) the spin excitations also correspond to the squeezed localized chain,
and (iii) the rules for adding charge (holon) and spin/anyon momenta {\it are identical}.
In addition, the numerical spectra agree very well with the sum of the spin (provided some renormalization
$\gamma$ of the energy scale, as shown in Fig.~\ref{fig:renormJ}) and charge spectra
(constructed independently) with the above-mentioned rule for momentum conservation.
We believe the small deviations can be attributed to finite size effects (which vanish when $J/t\rightarrow 0$).
Interestingly, the $J\rightarrow 0$ limit of the renormalization parameter $\gamma$,
$\gamma(0)\simeq 0.5515$ for $\rho=2/3$, is
independent on the anyon type (as the $J=0$ charge excitation spectrum).
In fact, in this limit, it should only depend on the probability of having two neighboring anyons.

\subsection{Anyonic and charge collective excitations}

Due to the above decoupling, the collective
anyonic and charge excitations can be deduced easily from the above
excitation spectra, by just applying selection rules.
Charge excitations occur between different charge branches at constant $m$
with energy transfer
\begin{equation}
{\cal E}_{p',p;m} ={\tilde E}_{\rm charge} (p', k_m) - {\tilde E}_{\rm charge} (p, k_m)\, ,
\end{equation}
and momentum transfer $K=K_{p'}-K_p$.
Anyonic excitations are characterized by
$\Delta p=0$ and are then given by
\begin{eqnarray}
{\cal E}_{p;m',m}
&=&E_{\rm anyon} (m') -  E_{\rm anyon} (m)
\nonumber \\
&+& {\tilde E}_{\rm charge} (p, k_{m'}) - {\tilde E}_{\rm charge} (p, k_m)\, ,
\label{eq:anyon_mode}
\end{eqnarray}
with momentum transfer $K=\rho(k_{m'}-k_m)$.
Note that the last two terms in Eq.~(\ref{eq:anyon_mode}) give a
finite size correction in the energy of order $1/L^2$.
In the thermodynamic limit, zero-energy anyonic excitations occur at momentum
$K_a=\rho k_a$ (and $2K_a$ if different), where
$k_a$ is the characteristic momentum (introduced
above) of the zero-energy mode of the pure chain. The location of both charge and anyonic
zero-energy modes are indicated in Fig.~\ref{fig:Maj}(b) for Ising anyons and in Figs.~\ref{fig:Fib}(a,b)
for Fibonacci anyons.

\subsection{Form of the eigenstates}

We now discuss briefly the structure of the eigenfunctions. Eq.~(2.14) of
Ref.~\onlinecite{ogata-shiba}
established that the ground-state of the $J\rightarrow 0$ limit of the electronic $t$-$J$ chain
can be written exactly as a product of a charge HCB wavefunction
times  a spin wavefunction identical to the ground-state of the 1D $S=1/2$ Heisenberg model.
Our results suggest that a similar product structure
might in fact also hold in the case of {\it all low-energy eigenstates} of 1D non-Abelian anyons at $J\ne 0$,
up to {\it finite size corrections}.
We speculate that the eigenfunctions can be approximately written as,
\begin{eqnarray}
\Psi_{p,m}(y_1,...,y_N; x_1,...,x_N)
&\simeq {\tilde \Phi}_{\rm charge}^{p} (y_1,...,y_N)\nonumber \\
&\times \,\chi_{\rm anyon}^{m}
(x_1,...,x_N)\, ,
\label{eq:sp-ch-wf}
\end{eqnarray}
where $y_j$ are the position of the (site) anyons on the $L$-site chain and
$x_i$ are the bond variables associated to them (see Fig.~\ref{fig:sketch}).
Here, ${\tilde \Phi}_{\rm charge}^{p}$ are the eigenstates (labelled by $p$) of
an {\it interacting} $L$-site HCB chain in the presence of a twist (i.e. flux)
$k_m$ in the boundary conditions and
$\chi_{\rm anyon}^{m}$ are the eigenstates (labelled by $m$) with momentum
$k_m$ of the interacting (undoped) anyonic chain of $L_a=\rho L$ sites.

\section{DMRG study of the model}

\begin{figure}
\includegraphics[width=\columnwidth]{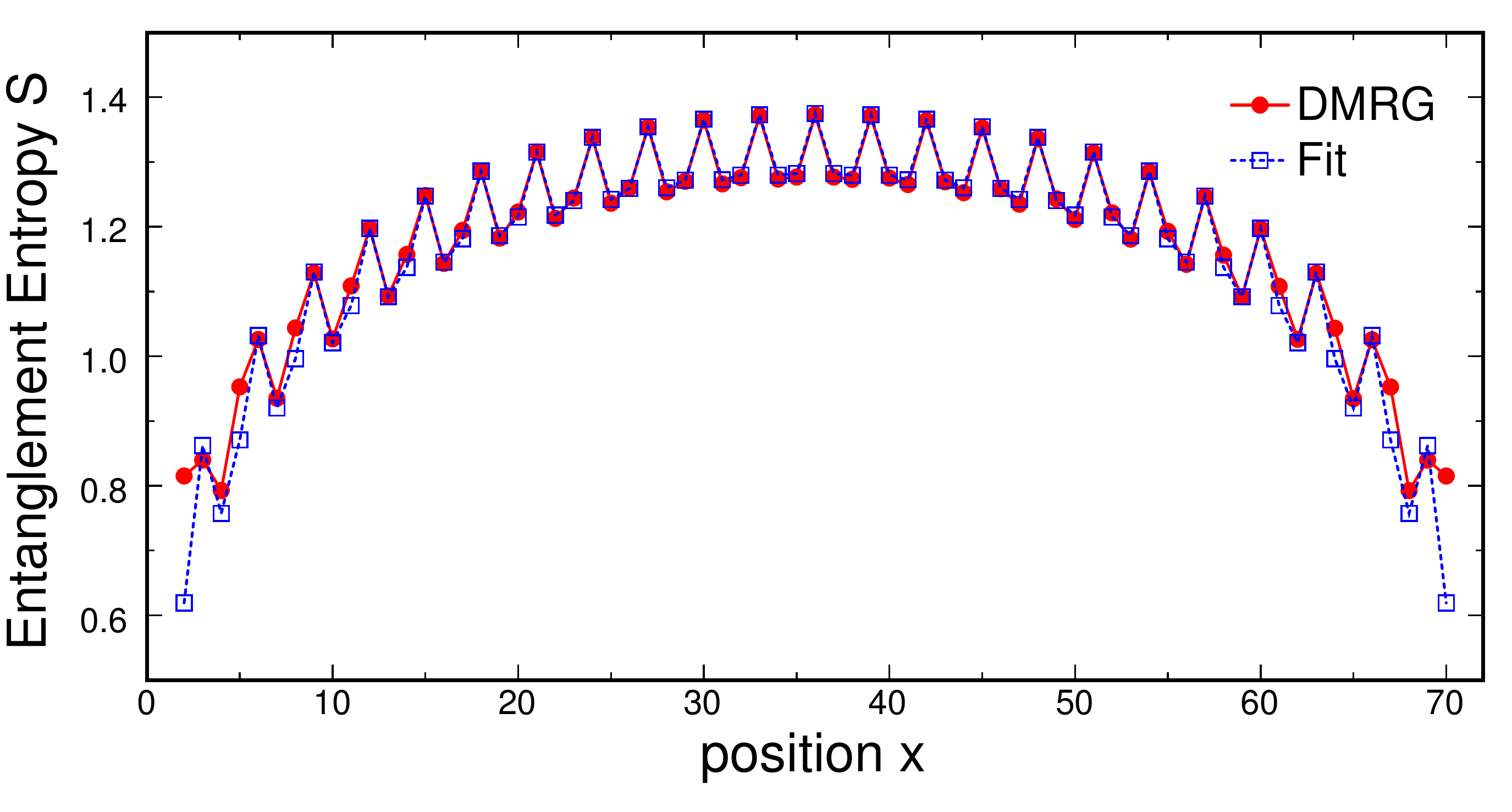}
\caption{Entanglement entropy obtained using DMRG for an open chain of length $L=72$, with $N=48$ Fibonacci anyons, and $J/t=0.3$. The EE predicted for a CFT with central charge $c=1.7$ is plotted and the agreement is seen to be excellent.}
\label{fig:entropy}
\end{figure}

We can use DMRG to compute the resulting CFT central charge
from the analysis of the von Neumann entanglement entropy (EE) of
an open chain (of length $L$) cut into two subsystems~\cite{Cardy2010}.
We focus here on the case of the diluted Fibonacci $t$-$J$ chain with $\rho=2/3$ and $J>0$, for which we expect the
anyonic part to be described by a $c=7/10$ tri-critical Ising CFT.
In Fig.~\ref{fig:entropy}, we plot the ground state's EE
\begin{equation}
S_A = S_B = -\text{Tr} \left[ \rho_A \log \rho_A \right]
\end{equation}
between subsystems $A$ and $B$, which are two connected segments of the open chain, as a function of the position of the cut along the chain. The calculation is somewhat non-standard (compared to usual spin systems) because the anyonic fusion tree bond variable $x_j$ labeling the $j$th link, which connects the two subsystems, i.e. the link across which one ``cuts'' the system in two, is shared by both subsystems of the chain. This shared link variable characterizes the overall topological charge of each subsystem. The reduced density matrices
\begin{equation}
\rho_A = \text{Tr}_{B} \rho, \qquad
\rho_B = \text{Tr}_{A} \rho
\end{equation}
of subsystems $A$ and $B$ are block diagonal with respect to this variable, i.e.
\begin{equation}
\rho_A = \bigoplus_{x_j} p_{x_j} \rho_{A,x_j}
,
\end{equation}
where $p_{x_j} = \text{Tr}[\Pi_{x_j} \rho] $ is the probability that the state will have topological charge $x_j$ on the $j$th link, and $\rho_{A,x_j} = \frac{1}{p_{x_j}} \text{Tr}_A [\Pi_{x_j} \rho]$ is the reduced density matrix for subsystem $A$ after projecting the $j$th link's variable onto the value $x_j$.

In order to verify that this Fibonacci $t$-$J$ chain results in a CFT with central charge $c=1.7$, we fit to the formula~\cite{Affleck91,Calabrese04,Laflorencie2006,Roux2009,Affleck2009,Cardy2010}
\begin{equation}
S(j)=a + b \, p_{x_j = \tau} + \frac{c}{6} \log{\left[L\sin{\left(\frac{\pi j}{L}\right)}\right]},
\end{equation}
where $a$ and $b$ are fitting parameters. The first term is a non-universal constant, which can include universal contributions, such as a boundary entropy. The second term is a phenomenologically motivated correction that is proportional to a local kinetic energy, i.e. $p_{x_j = \tau} = \langle n_{\text{link}} (j) \rangle$ where $n_{\text{link}}(j)$ is the density (occupation) operator of the $j$th link, and can include a contribution due to the boundary between the two subsystems. The third term is derived from CFT. We find the best fit for the parameter values $a= 0.31185$ and $b=-0.35547 $. As seen in Fig.~\ref{fig:entropy}, the agreement between the numerical results and the values provided by this expression is excellent.

\section{Conclusion and Outlook}

Motivated by the possible realization of non-Abelian Ising and Fibonacci quasiparticles in quantum Hall states and Majorana heterostructures and the importance of understanding their edge modes, we have investigated what happens if itinerant and interacting (charged) non-Abelian anyons are confined on a 1D chain, subject to a strong charging energy. Following a standard procedure for strongly correlated electronic systems, we have constructed simple low-energy effective models by truncating the Hilbert space to the relevant low-energy particles. Integrating out the high-energy virtual processes yields an ``exchange" interaction between anyons, which physically favors a particular fusion channel. The effective model generically takes the form of an anyonic
$t$-$J$ model, containing the exchange interaction ($J$) and the rate
($t$) of anyon ``hopping" between nearest-neighbor sites.
The central result of our work is that anyons fractionalize into their charge and (neutral) anyonic degrees
of freedom. This phenomenon closely resembles and generalizes the well-known spin-charge separation in
electronic Luttinger liquids. Incidentally, the numerically verification based on the identification of the many-body levels turned out to be more transparent for anyons, due to the absence of marginally irrelevant operators in the field theory description. The anyon fractionalization justifies
{\it a posteriori} the treatment of the edge theories of these topological phases as a direct product of the charge and neutral non-Abelian modes, even though the electric charge is not localized in current setups.

We note that the 1D electronic $t$-$J$ model exhibits an exact
supersymmetric point~\cite{bb90,tJ-1D-kawakami} at which
the full excitation spectrum can be obtained using the
Bethe ansatz~\cite{bbo91}.
It is left for future studies to investigate whether such an integrable point
also exists in 1D anyonic $t$-$J$ models.

Our simple description of interacting itinerant anyons now enables the investigation of realistic
setups for manipulating and/or braiding anyons for future quantum computation. It is also easy to extend this study to quasi-1D systems with more
than a single conduction channel: anyonic $t$-$J$ ``ladders'' could mimic such a case,
following the procedure for localized non-Abelian anyons~\cite{ladder-1}.
Whether fractionalization survives in two spatial dimensions is another important issue.
Localized anyons were shown to nucleate into a novel {\it gapped} quantum liquid~\cite{ladder-2} in two dimensions and a similar scenario might take place for itinerant anyons, with e.g. the anyonic degrees of freedom becoming gapped.

\section*{Acknowledgements}

DP acknowledges support by
the French Research Council (Agence Nationale de la
Recherche) under grant No. ANR 2010 BLAN 0406-01.
This work was granted access to the HPC resources of CALMIP under the allocation 2012-P1231. 
DP is also grateful to Nicolas Renon at CALMIP
(Toulouse, France) and to SGI (France) for support in the use of the Altix
SGI supercomputer.
AF acknowledges support by the NSF, Grant No. DMR-0955707.
PB and MT thank the Aspen Center for Physics for hospitality and
support under the NSF grant $\#PHY-1066293$.

\appendix

\section{Examples of Anyon Models}
\label{app:anyonmodels}

In this Appendix, we give detailed descriptions of the Ising, Fibonacci, and SU$(2)_{k}$ anyons models, and explain where they occur in non-Abelian quantum Hall states.

\subsection{Ising anyons}
  \label{sec:Ising}

The Ising anyon model is derived from the CFT that
describes the Ising model at criticality~\cite{Moore89b}. It is related to $\text{SU}(2)_{2}$ as its CFT can be obtained using the coset construction $\text{SU}(2)_{2}/\text{U}(1)_{4}$. It has topological charges $\mathcal{C}=\left\{I,\sigma,\psi \right\}$ (which respectively
correspond to vacuum, spin, and Majorana fermions in the CFT, and are sometimes denoted $0$, $\frac{1}{2}$, and $1$, because of the relation with $\text{SU}(2)_{2}$). The anyon model is described by (listing only the non-trivial $F$-symbols and $R$-symbols, i.e. those not listed are equal to one if their vertices are permitted by fusion, and equal to zero if they are not permitted):%
\begin{equation*}
\begin{tabular}{|l|}
\hline
$\mathcal{C}=\left\{I,\sigma,\psi \right\}, \quad I\times a=a,\quad \sigma \times \sigma%
=I+\psi,$ \\
$\qquad \qquad \sigma \times \psi=\sigma,\quad \psi \times \psi=I$ \\ \hline
\qquad \qquad$\left[ F_{\sigma}^{\sigma \sigma \sigma}\right] _{ef}=
%\left[ F_{\sigma \sigma}^{\sigma \sigma}\right] _{ef}=
\left[
\begin{array}{rr}
\frac{1}{\sqrt{2}} & \frac{1}{\sqrt{2}} \\
\frac{1}{\sqrt{2}} & \frac{-1}{\sqrt{2}}%
\end{array}\right] _{ef}^{\phantom{T}}$ \\
\qquad \qquad $\left[ F_{\psi}^{\sigma \psi \sigma}\right] _{\sigma \sigma}=%
\left[ F_{\sigma}^{\psi \sigma \psi}\right] _{\sigma \sigma_{\phantom{j}}}\!\!=
%\left[ F_{\psi \sigma}^{\sigma \psi}\right] _{\sigma \sigma}=
%\left[ F_{\sigma \psi}^{\psi \sigma}\right] _{\sigma \sigma}=
-1 $ \\ \hline
\qquad $R_{I}^{\sigma \sigma}=e^{-i\frac{\pi }{8}},\quad R_{\psi}^{\sigma \sigma}=e^{i\frac{3\pi }{8}},$ \\
\qquad $R_{\sigma}^{\sigma \psi}=R_{\sigma}^{\psi \sigma}=e^{-i\frac{\pi }{2}},\quad R_{I}^{\psi \psi}=-1$ \\ \hline
\qquad $d_{I}=d_{\psi}=1,\quad d_{\sigma_{\phantom{j}}}\!\!=\sqrt{2}, \quad \mathcal{D}=2$ \\ \hline
\qquad $\theta _{I}=1,\quad \theta_{\sigma}=e^{i\frac{\pi }{8}},\quad \theta _{\psi}=-1$ \\ \hline
\end{tabular}%
\end{equation*}%
where $e,f\in \left\{ I,\psi\right\} $.

\subsection{Fibonacci anyons}
  \label{sec:Fib}

The Fibonacci anyon model (also known as $\text{SO}(3)_{3}$, since it may be obtained from the $\text{SU}\left( 2\right) _{3}$ anyon model by restricting to integer spins $j=0,1$, though $\text{SO}\left( 3\right) _{k}$ is only allowed for $k=0~{\rm mod}~4$; as a Chern-Simons or WZW theory, it may, more properly, be equated with $\left( \text{G}_{2}\right)_{1}$) is known to be universal for TQC~\cite{Freedman02b}. It has two topological charges $\mathcal{C}=\left\{I,\tau \right\} $ (sometimes denoted $0$ and $1$, respectively, because of the relation with $\text{SU}\left( 2\right) _{3}$) and is described by (listing only the non-trivial $F$-symbols and $R$-symbols):%
\begin{equation*}
\begin{tabular}{|l|l|}
\hline
\multicolumn{2}{|l|}{$\mathcal{C}=\left\{I,\tau \right\}, \quad I\times I=I,\quad I\times \tau=\tau,\quad \tau\times \tau=I+\tau$} \\
\hline
\multicolumn{2}{|l|}{\qquad\qquad$\left[ F_{\tau}^{\tau \tau \tau}\right] _{ef}
%=\left[ F_{\tau \tau}^{\tau \tau}\right] _{ef}
=\left[
\begin{array}{cc}
\phi^{-1} & \phi^{-1/2} \\
\phi^{-1/2} & -\phi^{-1}%
\end{array}%
\right] _{ef_{\phantom{j}}}^{\phantom{T}}$} \\ \hline
\multicolumn{2}{|l|}{\qquad\qquad$R_{I}^{\tau \tau}=e^{-i4\pi /5},\quad R_{\tau}^{\tau \tau}=e^{i3\pi /5}$} \\ \hline
$d_{I}=1,\quad d_{\tau}=\phi,\quad \mathcal{D}=\sqrt{\phi+2}$ & $\theta _{I}=1,\quad \theta _{\tau}=e^{i\frac{%
4\pi }{5}}$ \\ \hline
\end{tabular}%
\end{equation*}%
where $\phi =\frac{1+\sqrt{5}}{2}$ is the Golden ratio.

\subsection{$\text{SU}\left(2\right)_{k}$}
  \label{sec:SU(2)_k}
The SU$\left( 2\right) _{k}$ anyon models (for $k$ an integer) are ``$q$-deformed'' versions of the usual SU$\left( 2\right) $ for
$q=e^{i\frac{2\pi }{k+2}}$, which, roughly speaking, means integers $n$ are replaced by $\left[ n\right]_{q}\equiv \frac{q^{n/2}-q^{-n/2}}{q^{1/2}-q^{-1/2}}$. These describe SU$\left(2\right) _{k}$ Chern-Simons theories~\cite{Witten89} and WZW CFTs~\cite{Wess71,Witten83}, and give rise to the Jones polynomials of knot theory~\cite{Jones85}. Their braiding statistics are known to be universal for TQC~\cite{Freedman02a} all $k$, except $k=1$, $2$, and $4$. They are described by:
\begin{widetext}
\begin{equation*}
\begin{tabular}{|l|l|}
\hline
\multicolumn{2}{|l|}{$\mathcal{C}=\left\{ 0,\frac{1}{2},\ldots ,\frac{k}{2}\right\}, \quad j_{1}\times j_{2}=\sum\limits_{j=\left|
j_{1}-j_{2}\right| }^{\min \left\{ j_{1}+j_{2},k-j_{1}-j_{2}\right\} }j$} \\
\hline
\multicolumn{2}{|l|}{$\left[ F_{j}^{j_{1},j_{2},j_{3}}\right]
_{j_{12},j_{23}}=\left( -1\right) ^{j_{1}+j_{2}+j_{3}+j}\sqrt{\left[
2j_{12}+1\right] _{q}\left[ 2j_{23}+1\right] _{q}}\left\{
\begin{array}{ccc}
j_{1} & j_{2} & j_{12} \\
j_{3} & j & j_{23}%
\end{array}%
\right\} _{q}^{\phantom{T}},$} \\
\multicolumn{2}{|l|}{$\left\{
\begin{array}{ccc}
j_{1} & j_{2} & j_{12} \\
j_{3} & j & j_{23}%
\end{array}%
\right\} _{q}=\Delta \left( j_{1},j_{2},j_{12}\right) \Delta \left(
j_{12},j_{3},j\right) \Delta \left( j_{2},j_{3},j_{23}\right) \Delta \left(
j_{1},j_{23},j\right) $} \\
\multicolumn{2}{|l|}{$\quad \quad \quad \quad \quad \quad \quad \quad \times
\sum\limits_{z}\left\{ \frac{\left( -1\right) ^{z}\left[ z+1\right] _{q}!}{%
\left[ z-j_{1}-j_{2}-j_{12}\right] _{q}!\left[ z-j_{12}-j_{3}-j\right] _{q}!%
\left[ z-j_{2}-j_{3}-j_{23}\right] _{q}!\left[ z-j_{1}-j_{23}-j\right] _{q}!}%
\right. $} \\
\multicolumn{2}{|l|}{$\quad \quad \quad \quad \quad \quad \quad \quad \quad
\quad \quad \times \left. \frac{1}{\left[ j_{1}+j_{2}+j_{3}+j-z\right] _{q}!%
\left[ j_{1}+j_{12}+j_{3}+j_{23}-z\right] _{q}!\left[ j_{2}+j_{12}+j+j_{23}-z%
\right] _{q}!}\right\}_{\phantom{g}} ,$} \\
\multicolumn{2}{|l|}{$\Delta \left( j_{1},j_{2},j_{3}\right) =\sqrt{\frac{%
\left[ -j_{1}+j_{2}+j_{3}\right] _{q}!\left[ j_{1}-j_{2}+j_{3}\right] _{q}!%
\left[ j_{1}+j_{2}-j_{3}\right] _{q}!}{\left[ j_{1}+j_{2}+j_{3}+1\right]
_{q}!}}^{\phantom{T}}_{\phantom{g}},\quad \quad \quad \left[ n\right] _{q}! \equiv \prod\limits_{m=1}^{n}%
\left[ m\right] _{q}$} \\ \hline
\multicolumn{2}{|l|}{$R_{j}^{j_{1},j_{2}}=\left( -1\right)
^{j-j_{1}-j_{2}}q^{\frac{1}{2}\left( j\left( j+1\right) -j_{1}\left(
j_{1}+1\right) -j_{2}\left( j_{2}+1\right) \right) }$} \\ \hline
$d_{j}= \left[ 2j+1 \right]_{q} =\frac{\sin \left( \frac{\left( 2j+1\right) \pi }{k+2}\right)^{\phantom{T}} }{\sin
\left( \frac{\pi }{k+2}\right)_{\phantom{g}} }, \quad \mathcal{D}=\frac{\sqrt{ \frac{k+2}{2}}}{\sin \left( \frac{\pi}{k+2} \right)_{\phantom{g}}}$ & $\theta _{j}=e^{i2\pi \frac{j\left(
j+1\right) }{k+2}}$ \\ \hline
\end{tabular}%
\end{equation*}
\end{widetext}
where $\left\{ \quad \right\} _{q}$ is a ``$q$-deformed'' version of the usual SU$\left( 2\right) $ $6j$-symbols (which correspond to $q=1$), and have been calculated in Ref.~\onlinecite{kr88} (see also Ref.~\onlinecite{as10}, for an introduction on how to calculate the $F$-symbols and an implementation in Mathematica). The sum in the definition of the $q$-deformed $6j$-symbol is over all integers in the range $\max\{ j_1 + j_2 + j_{12} ; j_{12} + j_{3} + j ; j_2 + j_3 + j_{23} ; j_1 + j_{23} + j  \} \leq z \leq \min\{ j_1 + j_2 + j_3 + j ; j_1 + j_{12} + j_3 + j_{23} ; j_2 + j_{12} + j + j_{23} \}$.

\subsection{Moore-Read, anti-Pfaffian, and Bonderson-Slingerland Hierarchy States}

The $\nu=1/m$ MR states~\cite{Moore91} are described by a spectrum restriction of the product of the Ising CFT with an Abelian $\text{U}\left( 1 \right)$. Specifically, the anyon model is
\begin{equation}
\text{MR} = \left. \text{Ising} \times \text{U}\left( 1 \right)_{m} \right|_{\mathcal{C}}
\end{equation}
where the restriction to the anyonic charge spectrum $\mathcal{C}$ is such that $I$ and $\psi$ Ising charges are paired with integer $\text{U}\left( 1 \right)$ fluxes, while $\sigma$ Ising charges are paired with half-integer $\text{U}\left( 1 \right)$ fluxes. The fundamental quasihole of the MR state has electric charge $e/2m$ (where the particle carries charge $-e$) and carries Ising topological charge $\sigma$. The $\nu=1/2$ MR state is a leading candidate for the experimentally observed $\nu=5/2$ and $7/2$ quantum Hall plateaus.

Taking the particle-hole conjugate of the MR state yields the aPf state~\cite{Lee07,Levin07}, which is another leading candidate for the $\nu=5/2$ and $7/2$ quantum Hall plateaus. The anyon model for the aPf state is simply obtained by taking the complex conjugate of the MR state's anyon model.

BS hierarchical states~\cite{Bonderson07d} may be obtained from the MR and aPf states by applying a hierarchical (or, equivalently, a composite fermion) construction to the $\text{U}\left( 1 \right)$ sector. The states built on MR may be written as
\begin{equation}
\text{BS}_{K} = \left. \text{Ising} \times \text{U}\left( 1 \right)_{K} \right|_{\mathcal{C}}
\end{equation}
where the $K$-matrix is determined by the details of the hierarchical construction over MR, and the spectrum restriction is similar to before. This produces Ising-type candidate states for all other observed second Landau level FQH filling fractions (including those observed at $\nu = 7/3$, $12/5$, $8/3$, and $14/5$). The quasiparticle excitation spectra of the BS states include excitations that carry the $\sigma$ Ising topological charge, but these are generally not the unique quasiparticle carrying the minimal electric charge.

\subsection{$k=3$ Read--Rezayi and NASS}

The particle-hole conjugate of the $k=3$, $M=1$ RR state~\cite{Read99} is a candidate for $\nu=12/5$, which is constructed from  the $\mathbb{Z}_{3}$-Parafermion (Pf$_{3}$) CFT and an Abelian $\text{U}\left( 1 \right)$. The braiding statistics of this state is described by the direct product of anyon models
\begin{equation}
\overline{\text{RR}}_{k=3,M=1} = \overline{ \text{Pf}_{3} \times \text{U}\left(1\right) } = \overline{\text{Fib}} \times \mathbb{Z}_{10}^{\left(3\right)}
,
\end{equation}
where the overline indicates complex conjugation and $\mathbb{Z}_{10}^{\left(3\right)}$ is an Abelian anyon model (using the notation of Ref.~\onlinecite{Bonderson07b,Bonderson07c}). The fundamental quasiholes of this state have electric charge $e/5$ and Fibonacci topological charge $\tau$.

The $k=2$, $M=1$ NASS state~\cite{Ardonne99}, based on $\text{SU}\left(3\right)_{k}$-parafermions, is a candidate for $\nu = 4/7$. Its braiding statistics is described by
\begin{equation}
\text{NASS}_{k=2,M=1} = \overline{\text{Fib}} \times \text{D}^{\prime} \left( \mathbb {Z}_{2} \right) \times \text{U}\left( 1 \right) \times \text{U}\left( 1 \right)
,
\end{equation}
where $\text{D}^{\prime} \left( \mathbb {Z}_{2} \right)$ is an Abelian theory similar to $\text{D} \left( \mathbb {Z}_{2} \right)$, the quantum double of $\mathbb {Z}_{2}$ (a.k.a. the toric code).
The two $U(1)$ factors describe the charge and spin of the particles.
Its data is listed in Ref.~\onlinecite{Bonderson07b} and also as $\nu=8$ in Table~2 of Ref.~\onlinecite{Kitaev06a}. The fundamental quasiholes of this state carrying Fib topological charge $\tau$, and electric charge of either $e/7$ or $2e/7$.

As these theories are the direct product of a Fibonacci theory with Abelian sectors, the braiding statistics of quasiparticle excitations carrying the non-trivial Fibonacci charge are computationally universal.

\section{Limiting the number of particles}
\label{app:truncation}

In this Appendix, we will provide some details on how we can restrict the
quasiparticle spectrum to just two quasiparticles, namely the trivial ``vacuum'' quasiparticle $I$, and
a ``fundamental'' or ``elementary'' excitation. This excitation has the smallest possible electric charge, and
is fundamental in the sense that all other excitations can be obtained from it by
repeated fusion. The reason behind this truncation is to come up with a model
which is tractable, because the Hilbert space grows exponentially in the number
of quasiparticle types. Apart from the truncation of the spectrum to two quasiparticle types only, we
also need that the energy associated with these particles is the same. To achieve
both goals, one has two tools, in principle. In particular, one can consider using a gate
which couples linearly to the charge, or consider the charging energy associated with
localizing the anyons on quantum dots. As we will show below, by using a gate alone,
one can arrange the system to have degenerate levels for the quasiparticles, but it turns out
that this lead to a degeneracy larger than two. To split this degeneracy, the quantum dot
charging is essential.

After we explain how we can restrict the number of
quasiparticles to these two types, we explain how we can map the obtained model to
the anyonic models introduced in Ref.~\onlinecite{ftl07} (in the case $\rho=1$).
We do not consider the MR and $k=3$ RR cases separately, but
directly consider the general case of the fermionic RR states for arbitrary $k$ (which includes MR at $k=2$). In addition, we only focus on those aspects
which we will need for our purposes in this paper.

We start by decomposing the operators creating the different types of particles into
two pieces, one associated with the non-Abelian statistics, the other with the electric charge
of the quasiparticles. The operator describing the fundamental quasiparticles, which have charge
$e/(k+2)$ is of the from
$\Phi^{1}_{1} e^{i \varphi /(\sqrt{k(k+2)})}$, where $\Phi^{1}_{1}$ is a parafermion field,
corresponding to $\sigma$ and $\sigma_1$ for $k=2$ and $k=3$, respectively, using
the notation of section \ref{subsec:MRRR}. The vertex operator $e^{i \varphi /(\sqrt{k(k+2)})}$, where $\varphi$ is a chiral bosonic scalar field,
gives the charge of the quasiparticle.

The energy of the quasiparticles in a finite geometry (such as the quantum dots used to localize
the quasiparticles) are proportional to the scaling dimensions of the fields creating the particles.
For each possible charge of the quasiparticles, $m \frac{e}{k+2}$, with $m=0,1,\ldots,k$, we only
consider the particles with the lowest scaling dimension. These are given by
$\Phi^{m}_{m} e^{i m \varphi /(\sqrt{k(k+2)})}$.
There are two contributions to the scaling dimension: the parafermion field
$\Phi^{m}_{m}$ contributes $\Delta_{\Phi^m_m} = \frac{m(k-m)}{k(k+2)}$ and the
charge sector contributes $\Delta_{\phi} (m) = \frac{m^2}{k(k+2)}$, giving a total scaling dimension
$\Delta_j = \frac{m}{(k+2)}$, which is therefore proportional to the charge of the excitations.

We would like to create a situation in which we have one non-trivial quasiparticle that is degenerate with
the vacuum, and an appreciable gap to the other types of quasiparticle excitations.
The first thing we could try to do is to lower the energy of the charge $e/(k+2)$ fundamental quasiparticle
by means of an added potential, such that it becomes degenerate with the
vacuum. However, we just saw that adding such a potential will actually create a set of
$k+1$ degenerate states. To circumvent this problem, we will assume that, in addition to
such a potential, there is also a charging energy proportional to $q^2$, where
$q$ is the charge of the excitation. Effectively, this modifies the amplitude of the quadratic
contribution to the scaling dimension, coming from the charge part~\cite{Bonderson10b}. Thus, by adding the
charging energy, and the energy associated with a suitable potential,
we indeed can create the situation of two degenerate lowest lying states
(one being the vacuum), separated from the others by a gap.

We have just argued that we can consider a chain of itinerant anyons, consisting of vacancies
with quantum numbers $(I,0)$ and fundamental quasiaparticle excitations with quantum numbers $(\Phi^1_1, e/(k+2))$.
Under fusion, the electric charge is merely additive, and we will therefore concentrate on the
non-Abelian sector only. In the original anyonic chain models, the constituent anyons
belong to the pure SU$(2)_k$ theory. The anyonic systems we study can be maped to these by noting that the $\Phi_{1}^{1}$ parafermionic field carries spin $j=1/2$ SU$(2)_k$ topological charge, together with some Abelian topological charges.

\bibliography{corr}

\begin{thebibliography}{75}
\expandafter\ifx\csname natexlab\endcsname\relax\def\natexlab#1{#1}\fi
\expandafter\ifx\csname bibnamefont\endcsname\relax
  \def\bibnamefont#1{#1}\fi
\expandafter\ifx\csname bibfnamefont\endcsname\relax
  \def\bibfnamefont#1{#1}\fi
\expandafter\ifx\csname citenamefont\endcsname\relax
  \def\citenamefont#1{#1}\fi
\expandafter\ifx\csname url\endcsname\relax
  \def\url#1{\texttt{#1}}\fi
\expandafter\ifx\csname urlprefix\endcsname\relax\def\urlprefix{URL }\fi
\providecommand{\bibinfo}[2]{#2}
\providecommand{\eprint}[2][]{\url{#2}}

\bibitem[{\citenamefont{Leinaas and Myrheim}(1977)}]{Leinaas77}
\bibinfo{author}{\bibfnamefont{J.~M.} \bibnamefont{Leinaas}} \bibnamefont{and}
  \bibinfo{author}{\bibfnamefont{J.}~\bibnamefont{Myrheim}},
  \bibinfo{journal}{Nuovo Cimento B} \textbf{\bibinfo{volume}{37B}},
  \bibinfo{pages}{1} (\bibinfo{year}{1977}).

\bibitem[{\citenamefont{Goldin et~al.}(1985)\citenamefont{Goldin, Menikoff, and
  Sharp}}]{Goldin85}
\bibinfo{author}{\bibfnamefont{G.~A.} \bibnamefont{Goldin}},
  \bibinfo{author}{\bibfnamefont{R.}~\bibnamefont{Menikoff}}, \bibnamefont{and}
  \bibinfo{author}{\bibfnamefont{D.~H.} \bibnamefont{Sharp}},
  \bibinfo{journal}{Phys. Rev. Lett.} \textbf{\bibinfo{volume}{54}},
  \bibinfo{pages}{603} (\bibinfo{year}{1985}).

\bibitem[{\citenamefont{Fredenhagen et~al.}(1989)\citenamefont{Fredenhagen,
  Rehren, and Schroer}}]{Fredenhagen89}
\bibinfo{author}{\bibfnamefont{K.}~\bibnamefont{Fredenhagen}},
  \bibinfo{author}{\bibfnamefont{K.~H.} \bibnamefont{Rehren}},
  \bibnamefont{and} \bibinfo{author}{\bibfnamefont{B.}~\bibnamefont{Schroer}},
  \bibinfo{journal}{Commun. Math. Phys.} \textbf{\bibinfo{volume}{125}},
  \bibinfo{pages}{201} (\bibinfo{year}{1989}).

\bibitem[{\citenamefont{Fr\"{o}hlich and Gabbiani}(1990)}]{Froehlich90}
\bibinfo{author}{\bibfnamefont{J.}~\bibnamefont{Fr\"{o}hlich}}
  \bibnamefont{and} \bibinfo{author}{\bibfnamefont{F.}~\bibnamefont{Gabbiani}},
  \bibinfo{journal}{Rev. Math. Phys.} \textbf{\bibinfo{volume}{2}},
  \bibinfo{pages}{251} (\bibinfo{year}{1990}).

\bibitem[{\citenamefont{Moore and Read}(1991)}]{Moore91}
\bibinfo{author}{\bibfnamefont{G.}~\bibnamefont{Moore}} \bibnamefont{and}
  \bibinfo{author}{\bibfnamefont{N.}~\bibnamefont{Read}},
  \bibinfo{journal}{Nucl. Phys. B} \textbf{\bibinfo{volume}{360}},
  \bibinfo{pages}{362} (\bibinfo{year}{1991}).

\bibitem[{\citenamefont{Read and Rezayi}(1999)}]{Read99}
\bibinfo{author}{\bibfnamefont{N.}~\bibnamefont{Read}} \bibnamefont{and}
  \bibinfo{author}{\bibfnamefont{E.}~\bibnamefont{Rezayi}},
  \bibinfo{journal}{Phys. Rev. B} \textbf{\bibinfo{volume}{59}},
  \bibinfo{pages}{8084} (\bibinfo{year}{1999}), \eprint{cond-mat/9809384}.

\bibitem[{\citenamefont{Lee et~al.}(2007)\citenamefont{Lee, Ryu, Nayak, and
  Fisher}}]{Lee07}
\bibinfo{author}{\bibfnamefont{S.-S.} \bibnamefont{Lee}},
  \bibinfo{author}{\bibfnamefont{S.}~\bibnamefont{Ryu}},
  \bibinfo{author}{\bibfnamefont{C.}~\bibnamefont{Nayak}}, \bibnamefont{and}
  \bibinfo{author}{\bibfnamefont{M.~P.~A.} \bibnamefont{Fisher}},
  \bibinfo{journal}{Phys. Rev. Lett.} \textbf{\bibinfo{volume}{99}},
  \bibinfo{pages}{236807} (\bibinfo{year}{2007}), \eprint{arXiv:0707.0478}.

\bibitem[{\citenamefont{Levin et~al.}(2007)\citenamefont{Levin, Halperin, and
  Rosenow}}]{Levin07}
\bibinfo{author}{\bibfnamefont{M.}~\bibnamefont{Levin}},
  \bibinfo{author}{\bibfnamefont{B.~I.} \bibnamefont{Halperin}},
  \bibnamefont{and} \bibinfo{author}{\bibfnamefont{B.}~\bibnamefont{Rosenow}},
  \bibinfo{journal}{Phys. Rev. Lett.} \textbf{\bibinfo{volume}{99}},
  \bibinfo{pages}{236806} (\bibinfo{year}{2007}), \eprint{arXiv:0707.0483}.

\bibitem[{\citenamefont{Bonderson and Slingerland}(2008)}]{Bonderson07d}
\bibinfo{author}{\bibfnamefont{P.}~\bibnamefont{Bonderson}} \bibnamefont{and}
  \bibinfo{author}{\bibfnamefont{J.~K.} \bibnamefont{Slingerland}},
  \bibinfo{journal}{Phys. Rev. B} \textbf{\bibinfo{volume}{78}},
  \bibinfo{pages}{067836} (\bibinfo{year}{2008}), \eprint{arXiv:0711.3204}.

\bibitem[{\citenamefont{Radu et~al.}(2008)\citenamefont{Radu, Miller, Marcus,
  Kastner, Pfeiffer, and West}}]{Radu08}
\bibinfo{author}{\bibfnamefont{I.~P.} \bibnamefont{Radu}},
  \bibinfo{author}{\bibfnamefont{J.~B.} \bibnamefont{Miller}},
  \bibinfo{author}{\bibfnamefont{C.~M.} \bibnamefont{Marcus}},
  \bibinfo{author}{\bibfnamefont{M.~A.} \bibnamefont{Kastner}},
  \bibinfo{author}{\bibfnamefont{L.~N.} \bibnamefont{Pfeiffer}},
  \bibnamefont{and} \bibinfo{author}{\bibfnamefont{K.~W.} \bibnamefont{West}},
  \bibinfo{journal}{Science} \textbf{\bibinfo{volume}{320}},
  \bibinfo{pages}{899} (\bibinfo{year}{2008}), \eprint{ar{X}iv:0803.3530}.

\bibitem[{\citenamefont{Willett et~al.}(2009)\citenamefont{Willett, Pfeiffer,
  and West}}]{Willett09a}
\bibinfo{author}{\bibfnamefont{R.~L.} \bibnamefont{Willett}},
  \bibinfo{author}{\bibfnamefont{L.~N.} \bibnamefont{Pfeiffer}},
  \bibnamefont{and} \bibinfo{author}{\bibfnamefont{K.~W.} \bibnamefont{West}},
  \bibinfo{journal}{Proc. Natl. Acad. Sci.} \textbf{\bibinfo{volume}{106}},
  \bibinfo{pages}{8853} (\bibinfo{year}{2009}), \eprint{arXiv:0807.0221}.

\bibitem[{\citenamefont{Willett et~al.}(2012)\citenamefont{Willett, Pfeiffer,
  and West}}]{Willett12}
\bibinfo{author}{\bibfnamefont{R.~L.} \bibnamefont{Willett}},
  \bibinfo{author}{\bibfnamefont{L.~N.} \bibnamefont{Pfeiffer}},
  \bibnamefont{and} \bibinfo{author}{\bibfnamefont{K.~W.} \bibnamefont{West}}
  (\bibinfo{year}{2012}), \eprint{arXiv:1204.1993}.

\bibitem[{\citenamefont{Willett et~al.}(1987)\citenamefont{Willett, Eisenstein,
  Stormer, Tsui, Gossard, and English}}]{Willett87}
\bibinfo{author}{\bibfnamefont{R.}~\bibnamefont{Willett}},
  \bibinfo{author}{\bibfnamefont{J.~P.} \bibnamefont{Eisenstein}},
  \bibinfo{author}{\bibfnamefont{H.~L.} \bibnamefont{Stormer}},
  \bibinfo{author}{\bibfnamefont{D.~C.} \bibnamefont{Tsui}},
  \bibinfo{author}{\bibfnamefont{A.~C.} \bibnamefont{Gossard}},
  \bibnamefont{and} \bibinfo{author}{\bibfnamefont{J.~H.}
  \bibnamefont{English}}, \bibinfo{journal}{Phys. Rev. Lett.}
  \textbf{\bibinfo{volume}{59}}, \bibinfo{pages}{1776} (\bibinfo{year}{1987}).

\bibitem[{\citenamefont{Pan et~al.}(1999)\citenamefont{Pan, Xia, Shvarts,
  Adams, Stormer, Tsui, Pfeiffer, Baldwin, and West}}]{Pan99}
\bibinfo{author}{\bibfnamefont{W.}~\bibnamefont{Pan}},
  \bibinfo{author}{\bibfnamefont{J.-S.} \bibnamefont{Xia}},
  \bibinfo{author}{\bibfnamefont{V.}~\bibnamefont{Shvarts}},
  \bibinfo{author}{\bibfnamefont{D.~E.} \bibnamefont{Adams}},
  \bibinfo{author}{\bibfnamefont{H.~L.} \bibnamefont{Stormer}},
  \bibinfo{author}{\bibfnamefont{D.~C.} \bibnamefont{Tsui}},
  \bibinfo{author}{\bibfnamefont{L.~N.} \bibnamefont{Pfeiffer}},
  \bibinfo{author}{\bibfnamefont{K.~W.} \bibnamefont{Baldwin}},
  \bibnamefont{and} \bibinfo{author}{\bibfnamefont{K.~W.} \bibnamefont{West}},
  \bibinfo{journal}{Phys. Rev. Lett.} \textbf{\bibinfo{volume}{83}},
  \bibinfo{pages}{3530} (\bibinfo{year}{1999}), \eprint{cond-mat/9907356}.

\bibitem[{\citenamefont{Eisenstein et~al.}(2002)\citenamefont{Eisenstein,
  Cooper, Pfeiffer, and West}}]{Eisenstein02}
\bibinfo{author}{\bibfnamefont{J.~P.} \bibnamefont{Eisenstein}},
  \bibinfo{author}{\bibfnamefont{K.~B.} \bibnamefont{Cooper}},
  \bibinfo{author}{\bibfnamefont{L.~N.} \bibnamefont{Pfeiffer}},
  \bibnamefont{and} \bibinfo{author}{\bibfnamefont{K.~W.} \bibnamefont{West}},
  \bibinfo{journal}{Phys. Rev. Lett.} \textbf{\bibinfo{volume}{88}},
  \bibinfo{pages}{076801} (\bibinfo{year}{2002}), \eprint{cond-mat/0110477}.

\bibitem[{\citenamefont{Xia et~al.}(2004)\citenamefont{Xia, Pan, Vicente,
  Adams, Sullivan, Stormer, Tsui, Pfeiffer, Baldwin, and West}}]{Xia04}
\bibinfo{author}{\bibfnamefont{J.~S.} \bibnamefont{Xia}},
  \bibinfo{author}{\bibfnamefont{W.}~\bibnamefont{Pan}},
  \bibinfo{author}{\bibfnamefont{C.~L.} \bibnamefont{Vicente}},
  \bibinfo{author}{\bibfnamefont{E.~D.} \bibnamefont{Adams}},
  \bibinfo{author}{\bibfnamefont{N.~S.} \bibnamefont{Sullivan}},
  \bibinfo{author}{\bibfnamefont{H.~L.} \bibnamefont{Stormer}},
  \bibinfo{author}{\bibfnamefont{D.~C.} \bibnamefont{Tsui}},
  \bibinfo{author}{\bibfnamefont{L.~N.} \bibnamefont{Pfeiffer}},
  \bibinfo{author}{\bibfnamefont{K.~W.} \bibnamefont{Baldwin}},
  \bibnamefont{and} \bibinfo{author}{\bibfnamefont{K.~W.} \bibnamefont{West}},
  \bibinfo{journal}{Phys. Rev. Lett.} \textbf{\bibinfo{volume}{93}},
  \bibinfo{pages}{176809} (\bibinfo{year}{2004}), \eprint{cond-mat/0406724}.

\bibitem[{\citenamefont{Kumar et~al.}(2010)\citenamefont{Kumar, Cs\'athy,
  Manfra, Pfeiffer, and West}}]{Kumar10}
\bibinfo{author}{\bibfnamefont{A.}~\bibnamefont{Kumar}},
  \bibinfo{author}{\bibfnamefont{G.~A.} \bibnamefont{Cs\'athy}},
  \bibinfo{author}{\bibfnamefont{M.~J.} \bibnamefont{Manfra}},
  \bibinfo{author}{\bibfnamefont{L.~N.} \bibnamefont{Pfeiffer}},
  \bibnamefont{and} \bibinfo{author}{\bibfnamefont{K.~W.} \bibnamefont{West}},
  \bibinfo{journal}{Phys. Rev. Lett.} \textbf{\bibinfo{volume}{105}},
  \bibinfo{pages}{246808} (\bibinfo{year}{2010}), \eprint{arXiv:1009.0237}.

\bibitem[{\citenamefont{Bonderson et~al.}(2012)\citenamefont{Bonderson,
  Feiguin, Moller, and Slingerland}}]{Bonderson09a}
\bibinfo{author}{\bibfnamefont{P.}~\bibnamefont{Bonderson}},
  \bibinfo{author}{\bibfnamefont{A.~E.} \bibnamefont{Feiguin}},
  \bibinfo{author}{\bibfnamefont{G.}~\bibnamefont{Moller}}, \bibnamefont{and}
  \bibinfo{author}{\bibfnamefont{J.~K.} \bibnamefont{Slingerland}},
  \bibinfo{journal}{Phys. Rev. Lett.} \textbf{\bibinfo{volume}{108}},
  \bibinfo{pages}{036806} (\bibinfo{year}{2012}), \eprint{arXiv:0901.4965}.

\bibitem[{\citenamefont{Hermanns}(2010)}]{h10}
\bibinfo{author}{\bibfnamefont{M.}~\bibnamefont{Hermanns}},
  \bibinfo{journal}{Phys. Rev. Lett.} \textbf{\bibinfo{volume}{104}},
  \bibinfo{pages}{056803} (\bibinfo{year}{2010}), \eprint{arXiv:0906.2073}.

\bibitem[{\citenamefont{Ardonne and Schoutens}(1999)}]{Ardonne99}
\bibinfo{author}{\bibfnamefont{E.}~\bibnamefont{Ardonne}} \bibnamefont{and}
  \bibinfo{author}{\bibfnamefont{K.}~\bibnamefont{Schoutens}},
  \bibinfo{journal}{Phys. Rev. Lett.} \textbf{\bibinfo{volume}{82}},
  \bibinfo{pages}{5096} (\bibinfo{year}{1999}), \eprint{cond-mat/9811352}.

\bibitem[{\citenamefont{{Volovik}}(1999)}]{Volovik99}
\bibinfo{author}{\bibfnamefont{G.~E.} \bibnamefont{{Volovik}}},
  \bibinfo{journal}{Soviet Journal of Experimental and Theoretical Physics
  Letters} \textbf{\bibinfo{volume}{70}}, \bibinfo{pages}{792}
  (\bibinfo{year}{1999}), \eprint{cond-mat/9911374}.

\bibitem[{\citenamefont{Read and Green}(2000)}]{Read00}
\bibinfo{author}{\bibfnamefont{N.}~\bibnamefont{Read}} \bibnamefont{and}
  \bibinfo{author}{\bibfnamefont{D.}~\bibnamefont{Green}},
  \bibinfo{journal}{Phys. Rev. B} \textbf{\bibinfo{volume}{61}},
  \bibinfo{pages}{10267} (\bibinfo{year}{2000}), \eprint{cond-mat/9906453}.

\bibitem[{\citenamefont{Kitaev}(2001)}]{Kitaev01a}
\bibinfo{author}{\bibfnamefont{A.~Y.} \bibnamefont{Kitaev}},
  \bibinfo{journal}{Physics-Uspekhi} \textbf{\bibinfo{volume}{44}},
  \bibinfo{pages}{131} (\bibinfo{year}{2001}), \eprint{cond-mat/0010440}.

\bibitem[{\citenamefont{Fu and Kane}(2008)}]{fk08}
\bibinfo{author}{\bibfnamefont{L.}~\bibnamefont{Fu}} \bibnamefont{and}
  \bibinfo{author}{\bibfnamefont{C.}~\bibnamefont{Kane}},
  \bibinfo{journal}{Phys. Rev. Lett.} \textbf{\bibinfo{volume}{100}},
  \bibinfo{pages}{096407} (\bibinfo{year}{2008}), \eprint{arXiv:0707.1692}.

\bibitem[{\citenamefont{Sau et~al.}(2010)\citenamefont{Sau, Lutchyn, Tewari,
  and Sarma}}]{slt10}
\bibinfo{author}{\bibfnamefont{J.}~\bibnamefont{Sau}},
  \bibinfo{author}{\bibfnamefont{R.}~\bibnamefont{Lutchyn}},
  \bibinfo{author}{\bibfnamefont{S.}~\bibnamefont{Tewari}}, \bibnamefont{and}
  \bibinfo{author}{\bibfnamefont{S.~D.} \bibnamefont{Sarma}},
  \bibinfo{journal}{Phys. Rev. Lett.} \textbf{\bibinfo{volume}{104}},
  \bibinfo{pages}{040502} (\bibinfo{year}{2010}), \eprint{arXiv:0907.2239}.

\bibitem[{\citenamefont{Alicea}(2010)}]{a10}
\bibinfo{author}{\bibfnamefont{J.}~\bibnamefont{Alicea}},
  \bibinfo{journal}{Phys. Rev. B} \textbf{\bibinfo{volume}{81}},
  \bibinfo{pages}{125318} (\bibinfo{year}{2010}), \eprint{arXiv:0912.2115}.

\bibitem[{\citenamefont{Lutchyn et~al.}(2010)\citenamefont{Lutchyn, Sau, and
  Das~Sarma}}]{Lutchyn10}
\bibinfo{author}{\bibfnamefont{R.~M.} \bibnamefont{Lutchyn}},
  \bibinfo{author}{\bibfnamefont{J.~D.} \bibnamefont{Sau}}, \bibnamefont{and}
  \bibinfo{author}{\bibfnamefont{S.}~\bibnamefont{Das~Sarma}},
  \bibinfo{journal}{Phys. Rev. Lett.} \textbf{\bibinfo{volume}{105}},
  \bibinfo{pages}{077001} (\bibinfo{year}{2010}), \eprint{arXiv:1002.4033}.

\bibitem[{\citenamefont{{Oreg} et~al.}(2010)\citenamefont{{Oreg}, {Refael}, and
  {von Oppen}}}]{Oreg10}
\bibinfo{author}{\bibfnamefont{Y.}~\bibnamefont{{Oreg}}},
  \bibinfo{author}{\bibfnamefont{G.}~\bibnamefont{{Refael}}}, \bibnamefont{and}
  \bibinfo{author}{\bibfnamefont{F.}~\bibnamefont{{von Oppen}}}
  (\bibinfo{year}{2010}), \eprint{arXiv:1003.1145}.

\bibitem[{\citenamefont{Alicea}(2012)}]{a12}
\bibinfo{author}{\bibfnamefont{J.}~\bibnamefont{Alicea}},
  \bibinfo{journal}{Rep. Prog. Phys.} \textbf{\bibinfo{volume}{75}},
  \bibinfo{pages}{076501} (\bibinfo{year}{2012}), \eprint{arXiv:1202.1293}.

\bibitem[{\citenamefont{Mourik et~al.}(2012)\citenamefont{Mourik, Zuo, Frolov,
  Plissard, Bakkers, and Kouwenhoven}}]{mzf12}
\bibinfo{author}{\bibfnamefont{V.}~\bibnamefont{Mourik}},
  \bibinfo{author}{\bibfnamefont{K.}~\bibnamefont{Zuo}},
  \bibinfo{author}{\bibfnamefont{S.}~\bibnamefont{Frolov}},
  \bibinfo{author}{\bibfnamefont{S.}~\bibnamefont{Plissard}},
  \bibinfo{author}{\bibfnamefont{E.}~\bibnamefont{Bakkers}}, \bibnamefont{and}
  \bibinfo{author}{\bibfnamefont{L.}~\bibnamefont{Kouwenhoven}},
  \bibinfo{journal}{Science} \textbf{\bibinfo{volume}{336}},
  \bibinfo{pages}{1003} (\bibinfo{year}{2012}), \eprint{arXiv:1204.2792}.

\bibitem[{\citenamefont{Rokhinson et~al.}(2012)\citenamefont{Rokhinson, Liu,
  and Furdyna}}]{Rokhinson12}
\bibinfo{author}{\bibfnamefont{L.~P.} \bibnamefont{Rokhinson}},
  \bibinfo{author}{\bibfnamefont{X.}~\bibnamefont{Liu}}, \bibnamefont{and}
  \bibinfo{author}{\bibfnamefont{J.~K.} \bibnamefont{Furdyna}}
  (\bibinfo{year}{2012}), \eprint{arXiv:1204.4212}.

\bibitem[{\citenamefont{Deng et~al.}(2012)\citenamefont{Deng, Yu, Huang,
  Larsson, Caroff, and Xu}}]{dyh12up}
\bibinfo{author}{\bibfnamefont{M.}~\bibnamefont{Deng}},
  \bibinfo{author}{\bibfnamefont{C.}~\bibnamefont{Yu}},
  \bibinfo{author}{\bibfnamefont{G.}~\bibnamefont{Huang}},
  \bibinfo{author}{\bibfnamefont{M.}~\bibnamefont{Larsson}},
  \bibinfo{author}{\bibfnamefont{P.}~\bibnamefont{Caroff}}, \bibnamefont{and}
  \bibinfo{author}{\bibfnamefont{H.}~\bibnamefont{Xu}} (\bibinfo{year}{2012}),
  \eprint{arXiv:1204.4130}.

\bibitem[{\citenamefont{Das et~al.}(2012)\citenamefont{Das, Ronen, Most, Oreg,
  Heiblum, and Shtrikman}}]{Das12}
\bibinfo{author}{\bibfnamefont{A.}~\bibnamefont{Das}},
  \bibinfo{author}{\bibfnamefont{Y.}~\bibnamefont{Ronen}},
  \bibinfo{author}{\bibfnamefont{Y.}~\bibnamefont{Most}},
  \bibinfo{author}{\bibfnamefont{Y.}~\bibnamefont{Oreg}},
  \bibinfo{author}{\bibfnamefont{M.}~\bibnamefont{Heiblum}}, \bibnamefont{and}
  \bibinfo{author}{\bibfnamefont{H.}~\bibnamefont{Shtrikman}}
  (\bibinfo{year}{2012}), \eprint{arXiv:1205.7073}.

\bibitem[{\citenamefont{Anderson}(1967)}]{anderson}
\bibinfo{author}{\bibfnamefont{P.~W.} \bibnamefont{Anderson}},
  \bibinfo{journal}{Phys. Rev.} \textbf{\bibinfo{volume}{164}},
  \bibinfo{pages}{352} (\bibinfo{year}{1967}).

\bibitem[{\citenamefont{Tomonaga}(1950)}]{t50}
\bibinfo{author}{\bibfnamefont{S.}~\bibnamefont{Tomonaga}},
  \bibinfo{journal}{Prog. Theor. Phys.} \textbf{\bibinfo{volume}{5}},
  \bibinfo{pages}{544} (\bibinfo{year}{1950}).

\bibitem[{\citenamefont{Luttinger}(1963)}]{l63}
\bibinfo{author}{\bibfnamefont{J.~M.} \bibnamefont{Luttinger}},
  \bibinfo{journal}{J. Math. Phys.} \textbf{\bibinfo{volume}{15}},
  \bibinfo{pages}{609} (\bibinfo{year}{1963}).

\bibitem[{\citenamefont{Haldane}(1981)}]{h81}
\bibinfo{author}{\bibfnamefont{F.~D.~M.} \bibnamefont{Haldane}},
  \bibinfo{journal}{J. Phys. C} \textbf{\bibinfo{volume}{14}},
  \bibinfo{pages}{2585} (\bibinfo{year}{1981}).

\bibitem[{\citenamefont{Poilblanc et~al.}(2012)\citenamefont{Poilblanc, Troyer,
  Ardonne, and Bonderson}}]{short-2012}
\bibinfo{author}{\bibfnamefont{D.}~\bibnamefont{Poilblanc}},
  \bibinfo{author}{\bibfnamefont{M.}~\bibnamefont{Troyer}},
  \bibinfo{author}{\bibfnamefont{E.}~\bibnamefont{Ardonne}}, \bibnamefont{and}
  \bibinfo{author}{\bibfnamefont{P.}~\bibnamefont{Bonderson}},
  \bibinfo{journal}{Phys. Rev. Lett.} \textbf{\bibinfo{volume}{108}},
  \bibinfo{pages}{207201} (\bibinfo{year}{2012}), \eprint{arXiv:1112.5950}.

\bibitem[{\citenamefont{Zhang and Rice}(1988)}]{ZhangRice}
\bibinfo{author}{\bibfnamefont{F.-C.} \bibnamefont{Zhang}} \bibnamefont{and}
  \bibinfo{author}{\bibfnamefont{T.~M.} \bibnamefont{Rice}},
  \bibinfo{journal}{Phys. Rev. B} \textbf{\bibinfo{volume}{37}},
  \bibinfo{pages}{3759} (\bibinfo{year}{1988}).

\bibitem[{\citenamefont{Hubbard}(1963)}]{Hubbard1}
\bibinfo{author}{\bibfnamefont{J.}~\bibnamefont{Hubbard}},
  \bibinfo{journal}{Proc. Roy. Soc. London A} \textbf{\bibinfo{volume}{276}},
  \bibinfo{pages}{238} (\bibinfo{year}{1963}).

\bibitem[{\citenamefont{Gutzwiller}(1963)}]{Hubbard2}
\bibinfo{author}{\bibfnamefont{M.~C.} \bibnamefont{Gutzwiller}},
  \bibinfo{journal}{Phys. Rev. Lett} \textbf{\bibinfo{volume}{10}},
  \bibinfo{pages}{159} (\bibinfo{year}{1963}).

\bibitem[{\citenamefont{Kanamori}(1963)}]{Hubbard3}
\bibinfo{author}{\bibfnamefont{J.}~\bibnamefont{Kanamori}},
  \bibinfo{journal}{Prog. of Theor. Phys. (Kyoto)}
  \textbf{\bibinfo{volume}{30}}, \bibinfo{pages}{275} (\bibinfo{year}{1963}).

\bibitem[{\citenamefont{Bares and Blatter}(1990)}]{bb90}
\bibinfo{author}{\bibfnamefont{P.~A.} \bibnamefont{Bares}} \bibnamefont{and}
  \bibinfo{author}{\bibfnamefont{G.}~\bibnamefont{Blatter}},
  \bibinfo{journal}{Phys. Rev. Lett.} \textbf{\bibinfo{volume}{64}},
  \bibinfo{pages}{2567} (\bibinfo{year}{1990}).

\bibitem[{\citenamefont{Kawakami and Yang}(1990)}]{tJ-1D-kawakami}
\bibinfo{author}{\bibfnamefont{N.}~\bibnamefont{Kawakami}} \bibnamefont{and}
  \bibinfo{author}{\bibfnamefont{S.-K.} \bibnamefont{Yang}},
  \bibinfo{journal}{Phys. Rev. Lett.} \textbf{\bibinfo{volume}{65}},
  \bibinfo{pages}{2309} (\bibinfo{year}{1990}).

\bibitem[{\citenamefont{Bares et~al.}(1991)\citenamefont{Bares, Blatter, and
  Ogata}}]{bbo91}
\bibinfo{author}{\bibfnamefont{P.-A.} \bibnamefont{Bares}},
  \bibinfo{author}{\bibfnamefont{G.}~\bibnamefont{Blatter}}, \bibnamefont{and}
  \bibinfo{author}{\bibfnamefont{M.}~\bibnamefont{Ogata}},
  \bibinfo{journal}{Phys. Rev. B} \textbf{\bibinfo{volume}{44}},
  \bibinfo{pages}{130} (\bibinfo{year}{1991}).

\bibitem[{\citenamefont{Ogata et~al.}(1991)\citenamefont{Ogata, Luchini,
  Sorella, and Assaad}}]{tJ-1D}
\bibinfo{author}{\bibfnamefont{M.}~\bibnamefont{Ogata}},
  \bibinfo{author}{\bibfnamefont{M.~U.} \bibnamefont{Luchini}},
  \bibinfo{author}{\bibfnamefont{S.}~\bibnamefont{Sorella}}, \bibnamefont{and}
  \bibinfo{author}{\bibfnamefont{F.~F.} \bibnamefont{Assaad}},
  \bibinfo{journal}{Phys. Rev. Lett.} \textbf{\bibinfo{volume}{66}},
  \bibinfo{pages}{2391} (\bibinfo{year}{1991}).

\bibitem[{\citenamefont{Bonderson}(2009)}]{Bonderson09b}
\bibinfo{author}{\bibfnamefont{P.}~\bibnamefont{Bonderson}},
  \bibinfo{journal}{Phys. Rev. Lett.} \textbf{\bibinfo{volume}{103}},
  \bibinfo{pages}{110403} (\bibinfo{year}{2009}), \eprint{arXiv:0905.2726}.

\bibitem[{\citenamefont{Bonderson et~al.}(2010)\citenamefont{Bonderson, Nayak,
  and Shtengel}}]{Bonderson10b}
\bibinfo{author}{\bibfnamefont{P.}~\bibnamefont{Bonderson}},
  \bibinfo{author}{\bibfnamefont{C.}~\bibnamefont{Nayak}}, \bibnamefont{and}
  \bibinfo{author}{\bibfnamefont{K.}~\bibnamefont{Shtengel}},
  \bibinfo{journal}{Phys. Rev. B} \textbf{\bibinfo{volume}{81}},
  \bibinfo{pages}{165308} (\bibinfo{year}{2010}), \eprint{arXiv:0909.1056}.

\bibitem[{\citenamefont{Messiah}(1962)}]{book:messiah62}
\bibinfo{author}{\bibfnamefont{A.}~\bibnamefont{Messiah}},
  \emph{\bibinfo{title}{Quantum Mechanics}}
  (\bibinfo{publisher}{North-Holland}, \bibinfo{address}{Amsterdam},
  \bibinfo{year}{1962}).

\bibitem[{\citenamefont{Feiguin et~al.}(2007)\citenamefont{Feiguin, Trebst,
  Ludwig, Troyer, Kitaev, Wang, and Freedman}}]{ftl07}
\bibinfo{author}{\bibfnamefont{A.}~\bibnamefont{Feiguin}},
  \bibinfo{author}{\bibfnamefont{S.}~\bibnamefont{Trebst}},
  \bibinfo{author}{\bibfnamefont{A.~W.~W.} \bibnamefont{Ludwig}},
  \bibinfo{author}{\bibfnamefont{M.}~\bibnamefont{Troyer}},
  \bibinfo{author}{\bibfnamefont{A.}~\bibnamefont{Kitaev}},
  \bibinfo{author}{\bibfnamefont{Z.}~\bibnamefont{Wang}}, \bibnamefont{and}
  \bibinfo{author}{\bibfnamefont{M.~H.} \bibnamefont{Freedman}},
  \bibinfo{journal}{Phys. Rev. Lett.} \textbf{\bibinfo{volume}{98}},
  \bibinfo{pages}{60409} (\bibinfo{year}{2007}), \eprint{cond-mat/0612341}.

\bibitem[{\citenamefont{Andrews et~al.}(1984)\citenamefont{Andrews, Baxter, and
  Forrester}}]{abf84}
\bibinfo{author}{\bibfnamefont{G.}~\bibnamefont{Andrews}},
  \bibinfo{author}{\bibfnamefont{R.}~\bibnamefont{Baxter}}, \bibnamefont{and}
  \bibinfo{author}{\bibfnamefont{P.}~\bibnamefont{Forrester}},
  \bibinfo{journal}{J. Stat. Phys.} \textbf{\bibinfo{volume}{35}},
  \bibinfo{pages}{193} (\bibinfo{year}{1984}).

\bibitem[{\citenamefont{Ogata and Shiba}(1990)}]{ogata-shiba}
\bibinfo{author}{\bibfnamefont{M.}~\bibnamefont{Ogata}} \bibnamefont{and}
  \bibinfo{author}{\bibfnamefont{H.}~\bibnamefont{Shiba}},
  \bibinfo{journal}{Phys. Rev. B} \textbf{\bibinfo{volume}{41}},
  \bibinfo{pages}{2326} (\bibinfo{year}{1990}).

\bibitem[{\citenamefont{Shiba and Ogata}(1991)}]{shiba-ogata}
\bibinfo{author}{\bibfnamefont{H.}~\bibnamefont{Shiba}} \bibnamefont{and}
  \bibinfo{author}{\bibfnamefont{M.}~\bibnamefont{Ogata}},
  \bibinfo{journal}{Int. J. Mod. Phys. B} \textbf{\bibinfo{volume}{5}},
  \bibinfo{pages}{31} (\bibinfo{year}{1991}).

\bibitem[{\citenamefont{Parola and Sorella}(1992)}]{parola-sorella-92}
\bibinfo{author}{\bibfnamefont{A.}~\bibnamefont{Parola}} \bibnamefont{and}
  \bibinfo{author}{\bibfnamefont{S.}~\bibnamefont{Sorella}},
  \bibinfo{journal}{Phys. Rev. B} \textbf{\bibinfo{volume}{45}},
  \bibinfo{pages}{13156} (\bibinfo{year}{1992}).

\bibitem[{\citenamefont{Cardy and Calabrese}(2010)}]{Cardy2010}
\bibinfo{author}{\bibfnamefont{J.}~\bibnamefont{Cardy}} \bibnamefont{and}
  \bibinfo{author}{\bibfnamefont{P.}~\bibnamefont{Calabrese}},
  \bibinfo{journal}{J. Stat. Mech.} p. \bibinfo{pages}{P04023}
  (\bibinfo{year}{2010}), \eprint{arXiv:1002.4353}.

\bibitem[{\citenamefont{Affleck and Ludwig}(1991)}]{Affleck91}
\bibinfo{author}{\bibfnamefont{I.}~\bibnamefont{Affleck}} \bibnamefont{and}
  \bibinfo{author}{\bibfnamefont{A.~W.~W.} \bibnamefont{Ludwig}},
  \bibinfo{journal}{Phys. Rev. Lett.} \textbf{\bibinfo{volume}{67}},
  \bibinfo{pages}{161} (\bibinfo{year}{1991}).

\bibitem[{\citenamefont{Calabrese and Cardy}(2004)}]{Calabrese04}
\bibinfo{author}{\bibfnamefont{P.}~\bibnamefont{Calabrese}} \bibnamefont{and}
  \bibinfo{author}{\bibfnamefont{J.}~\bibnamefont{Cardy}}, \bibinfo{journal}{J.
  Stat. Mech.} \textbf{\bibinfo{volume}{0406}}, \bibinfo{pages}{P06002}
  (\bibinfo{year}{2004}), \eprint{hep-th/0405152}.

\bibitem[{\citenamefont{Laflorencie et~al.}(2006)\citenamefont{Laflorencie,
  Sorensen, Chang, and Affleck}}]{Laflorencie2006}
\bibinfo{author}{\bibfnamefont{N.}~\bibnamefont{Laflorencie}},
  \bibinfo{author}{\bibfnamefont{E.~S.} \bibnamefont{Sorensen}},
  \bibinfo{author}{\bibfnamefont{M.-S.} \bibnamefont{Chang}}, \bibnamefont{and}
  \bibinfo{author}{\bibfnamefont{I.}~\bibnamefont{Affleck}},
  \bibinfo{journal}{Phys. Rev. Lett.} \textbf{\bibinfo{volume}{96}},
  \bibinfo{pages}{100603} (\bibinfo{year}{2006}), \eprint{cond-mat/0512475}.

\bibitem[{\citenamefont{Roux et~al.}(2009)\citenamefont{Roux, Capponi,
  Lecheminant, and Azaria}}]{Roux2009}
\bibinfo{author}{\bibfnamefont{G.}~\bibnamefont{Roux}},
  \bibinfo{author}{\bibfnamefont{S.}~\bibnamefont{Capponi}},
  \bibinfo{author}{\bibfnamefont{P.}~\bibnamefont{Lecheminant}},
  \bibnamefont{and} \bibinfo{author}{\bibfnamefont{P.}~\bibnamefont{Azaria}},
  \bibinfo{journal}{Eur. Phys. J. B} \textbf{\bibinfo{volume}{68}},
  \bibinfo{pages}{293} (\bibinfo{year}{2009}), \eprint{arXiv:0807.0412}.

\bibitem[{\citenamefont{Affleck et~al.}(2009)\citenamefont{Affleck,
  Laflorencie, and Sorensen}}]{Affleck2009}
\bibinfo{author}{\bibfnamefont{I.}~\bibnamefont{Affleck}},
  \bibinfo{author}{\bibfnamefont{N.}~\bibnamefont{Laflorencie}},
  \bibnamefont{and} \bibinfo{author}{\bibfnamefont{E.~S.}
  \bibnamefont{Sorensen}}, \bibinfo{journal}{J. Phys. A: Math. Theor.}
  \textbf{\bibinfo{volume}{42}}, \bibinfo{pages}{504009}
  (\bibinfo{year}{2009}), \eprint{arXiv:0906.1809}.

\bibitem[{\citenamefont{Poilblanc et~al.}(2011)\citenamefont{Poilblanc, Ludwig,
  Trebst, and Troyer}}]{ladder-1}
\bibinfo{author}{\bibfnamefont{D.}~\bibnamefont{Poilblanc}},
  \bibinfo{author}{\bibfnamefont{A.~W.} \bibnamefont{Ludwig}},
  \bibinfo{author}{\bibfnamefont{S.}~\bibnamefont{Trebst}}, \bibnamefont{and}
  \bibinfo{author}{\bibfnamefont{M.}~\bibnamefont{Troyer}},
  \bibinfo{journal}{Phys. Rev. B} \textbf{\bibinfo{volume}{83}},
  \bibinfo{pages}{134439} (\bibinfo{year}{2011}), \eprint{arXiv:1101.1186}.

\bibitem[{\citenamefont{Ludwig et~al.}(2011)\citenamefont{Ludwig, Poilblanc,
  Trebst, and Troyer}}]{ladder-2}
\bibinfo{author}{\bibfnamefont{A.~W.} \bibnamefont{Ludwig}},
  \bibinfo{author}{\bibfnamefont{D.}~\bibnamefont{Poilblanc}},
  \bibinfo{author}{\bibfnamefont{S.}~\bibnamefont{Trebst}}, \bibnamefont{and}
  \bibinfo{author}{\bibfnamefont{M.}~\bibnamefont{Troyer}},
  \bibinfo{journal}{New. J. Phys.} \textbf{\bibinfo{volume}{13}},
  \bibinfo{pages}{045014} (\bibinfo{year}{2011}), \eprint{arXiv:1003.3453}.

\bibitem[{\citenamefont{Moore and Seiberg}(1989)}]{Moore89b}
\bibinfo{author}{\bibfnamefont{G.}~\bibnamefont{Moore}} \bibnamefont{and}
  \bibinfo{author}{\bibfnamefont{N.}~\bibnamefont{Seiberg}},
  \bibinfo{journal}{Commun. Math. Phys.} \textbf{\bibinfo{volume}{123}},
  \bibinfo{pages}{177} (\bibinfo{year}{1989}).

\bibitem[{\citenamefont{Freedman
  et~al.}(2002{\natexlab{a}})\citenamefont{Freedman, Larsen, and
  Wang}}]{Freedman02b}
\bibinfo{author}{\bibfnamefont{M.~H.} \bibnamefont{Freedman}},
  \bibinfo{author}{\bibfnamefont{M.~J.} \bibnamefont{Larsen}},
  \bibnamefont{and} \bibinfo{author}{\bibfnamefont{Z.}~\bibnamefont{Wang}},
  \bibinfo{journal}{Commun. Math. Phys.} \textbf{\bibinfo{volume}{228}},
  \bibinfo{pages}{177} (\bibinfo{year}{2002}{\natexlab{a}}),
  \eprint{math/0103200}.

\bibitem[{\citenamefont{Witten}(1989)}]{Witten89}
\bibinfo{author}{\bibfnamefont{E.}~\bibnamefont{Witten}},
  \bibinfo{journal}{Comm. Math. Phys.} \textbf{\bibinfo{volume}{121}},
  \bibinfo{pages}{351} (\bibinfo{year}{1989}).

\bibitem[{\citenamefont{Wess and Zumino}(1971)}]{Wess71}
\bibinfo{author}{\bibfnamefont{J.}~\bibnamefont{Wess}} \bibnamefont{and}
  \bibinfo{author}{\bibfnamefont{B.}~\bibnamefont{Zumino}},
  \bibinfo{journal}{Phys. Lett. B} \textbf{\bibinfo{volume}{37}},
  \bibinfo{pages}{95} (\bibinfo{year}{1971}).

\bibitem[{\citenamefont{Witten}(1983)}]{Witten83}
\bibinfo{author}{\bibfnamefont{E.}~\bibnamefont{Witten}},
  \bibinfo{journal}{Nucl. Phys. B} \textbf{\bibinfo{volume}{223}},
  \bibinfo{pages}{422} (\bibinfo{year}{1983}).

\bibitem[{\citenamefont{Jones}(1985)}]{Jones85}
\bibinfo{author}{\bibfnamefont{V.~F.~R.} \bibnamefont{Jones}},
  \bibinfo{journal}{Bull. Am. Math. Soc.} \textbf{\bibinfo{volume}{12}},
  \bibinfo{pages}{103} (\bibinfo{year}{1985}).

\bibitem[{\citenamefont{Freedman
  et~al.}(2002{\natexlab{b}})\citenamefont{Freedman, Larsen, and
  Wang}}]{Freedman02a}
\bibinfo{author}{\bibfnamefont{M.~H.} \bibnamefont{Freedman}},
  \bibinfo{author}{\bibfnamefont{M.~J.} \bibnamefont{Larsen}},
  \bibnamefont{and} \bibinfo{author}{\bibfnamefont{Z.}~\bibnamefont{Wang}},
  \bibinfo{journal}{Commun. Math. Phys.} \textbf{\bibinfo{volume}{227}},
  \bibinfo{pages}{605} (\bibinfo{year}{2002}{\natexlab{b}}),
  \eprint{quant-ph/0001108}.

\bibitem[{\citenamefont{Kirillov and Reshetikhin}(1988)}]{kr88}
\bibinfo{author}{\bibfnamefont{A.}~\bibnamefont{Kirillov}} \bibnamefont{and}
  \bibinfo{author}{\bibfnamefont{N.}~\bibnamefont{Reshetikhin}}, in
  \emph{\bibinfo{booktitle}{Infinite dimensional Lie algebras and groups}},
  edited by \bibinfo{editor}{\bibfnamefont{V.~G.} \bibnamefont{Kac}}
  (\bibinfo{publisher}{World Scientific}, \bibinfo{address}{Singapore},
  \bibinfo{year}{1988}), p. \bibinfo{pages}{285}, \bibinfo{note}{proceedings of
  the conference held at CIRM, Luminy, Marseille}.

\bibitem[{\citenamefont{Ardonne and Slingerland}(2010)}]{as10}
\bibinfo{author}{\bibfnamefont{E.}~\bibnamefont{Ardonne}} \bibnamefont{and}
  \bibinfo{author}{\bibfnamefont{J.}~\bibnamefont{Slingerland}},
  \bibinfo{journal}{J. Phys. A} \textbf{\bibinfo{volume}{43}},
  \bibinfo{pages}{395205} (\bibinfo{year}{2010}), \eprint{arXiv:1004.5456}.

\bibitem[{\citenamefont{Bonderson}(2007)}]{Bonderson07b}
\bibinfo{author}{\bibfnamefont{P.~H.} \bibnamefont{Bonderson}}, Ph.D. thesis
  (\bibinfo{year}{2007}).

\bibitem[{\citenamefont{Bonderson et~al.}(2008)\citenamefont{Bonderson,
  Shtengel, and Slingerland}}]{Bonderson07c}
\bibinfo{author}{\bibfnamefont{P.}~\bibnamefont{Bonderson}},
  \bibinfo{author}{\bibfnamefont{K.}~\bibnamefont{Shtengel}}, \bibnamefont{and}
  \bibinfo{author}{\bibfnamefont{J.~K.} \bibnamefont{Slingerland}},
  \bibinfo{journal}{Annals of Physics} \textbf{\bibinfo{volume}{323}},
  \bibinfo{pages}{2709} (\bibinfo{year}{2008}), \eprint{arXiv:0707.4206}.

\bibitem[{\citenamefont{Kitaev}(2006)}]{Kitaev06a}
\bibinfo{author}{\bibfnamefont{A.}~\bibnamefont{Kitaev}},
  \bibinfo{journal}{Annals Phys.} \textbf{\bibinfo{volume}{321}},
  \bibinfo{pages}{2} (\bibinfo{year}{2006}), \eprint{cond-mat/0506438}.

\bibitem[{\citenamefont{Trebst et~al.}(2008)\citenamefont{Trebst, Troyer, Wang,
  and Ludwig}}]{Trebst08a}
\bibinfo{author}{\bibfnamefont{S.}~\bibnamefont{Trebst}},
  \bibinfo{author}{\bibfnamefont{M.}~\bibnamefont{Troyer}},
  \bibinfo{author}{\bibfnamefont{Z.}~\bibnamefont{Wang}}, \bibnamefont{and}
  \bibinfo{author}{\bibfnamefont{A.~W.~W.} \bibnamefont{Ludwig}},
  \bibinfo{journal}{Prog. Theor. Phys. Supp.} \textbf{\bibinfo{volume}{176}},
  \bibinfo{pages}{384} (\bibinfo{year}{2008}), \eprint{arXiv:0902.3275}.

\end{thebibliography}

%%%%%%%%%%%%%%%%%%%%%%%%%%%%%%%%%%%%%%%%%%%%%%%%%%%%%%%%%%%%%%%%%%%%%%%%%%%%%%%%
%% Bibliography
%%%%%%%%%%%%%%%%%%%%%%%%%%%%%%%%%%%%%%%%%%%%%%%%%%%%%%%%%%%%%%%%%%%%%%%%%%%%%%%%
%\begin{thebibliography}{99}
%
%% Topological order
%
%
%
%\end{thebibliography}

\end{document}